\newif\ifpdflatex    
\pdflatextrue           

\documentclass[usenatbib]{mn2e}
\usepackage{graphicx}
\usepackage{amsbsy}
\def\lesssim{\mathrel{\hbox{\rlap{\hbox{\lower5pt\hbox{$\sim$}}}\hbox{$<$}}}}
\def\gtrsim{\mathrel{\hbox{\rlap{\hbox{\lower5pt\hbox{$\sim$}}}\hbox{$>$}}}}
\title[A Failed Supernova?]{The search for failed supernovae with the Large Binocular Telescope: confirmation of a disappearing star}
\author[Adams et al.]
  {\parbox{18cm}{S.~M.~Adams$^{1,2}$, C.~S.~Kochanek$^{2,3}$, J.~R.~Gerke$^{2}$, K.~Z.~Stanek$^{2,3}$, and X. Dai$^{4}$}
  \\
  \\
  $^{1}$ Cahill Center for Astrophysics, California Institute of Technology, Pasadena, CA 91125, USA\\
  $^{2}$ Dept.\ of Astronomy, The Ohio State University, 140 W.\ 18th   Ave., Columbus, OH 43210, USA\\
  $^{3}$ Center for Cosmology and AstroParticle Physics (CCAPP), The Ohio State University, 191 W.\ Woodruff Ave., Columbus, OH 43210, USA\\
  $^{4}$ Department of Physics and Astronomy, University of Oklahoma, 440 W. Brooks St., Norman, OK 73019, USA \\
  E-mail: sma@astro.caltech.edu}
\begin{document}
\voffset -1.5cm
\maketitle

\begin{abstract}
We present \emph{Hubble Space Telescope} imaging confirming the optical disappearance of the failed supernova (SN) candidate identified by \citet{Gerke15}.  This $\sim$$25~M_{\odot}$ red supergiant experienced a weak $\sim$$10^{6}~L_{\odot}$ optical outburst in 2009 and is now at least 5 magnitudes fainter than the progenitor in the optical.  The mid-IR flux has slowly decreased to the lowest levels since the first measurements in 2004.  There is faint (2000-$3000~L_{\odot}$) near-IR emission likely associated with the source.  We find the late-time evolution of the source to be inconsistent with obscuration from an ejected, dusty shell.  Models of the spectral energy distribution indicate that the remaining bolometric luminosity is $>$6 times fainter than that of the progenitor and is decreasing as $\sim$$t^{-4/3}$.  We conclude that the transient is unlikely to be a SN impostor or stellar merger.  The event is consistent with the ejection of the envelope of a red supergiant in a failed SN and the late-time emission could be powered by fallback accretion on to a newly-formed black hole.  Future IR and X-ray observations are needed to confirm this interpretation of the fate for the star.
\\
\\
\end{abstract}

\begin{keywords}
stars: massive -- supernovae: individual (N6946-BH1) -- black hole physics
\end{keywords}

\section{Introduction}

Supernova (SN) surveys for the deaths of massive stars search for a sudden brightening of a source.  However, it is expected that some fraction of massive stars experience a failed SN, forming a black hole without a luminous SN.  While this idea is most widely accepted for very high mass stars at lower metallicity \citep{Heger03}, evidence has recently emerged suggesting that failed SN may also occur in red supergiants (RSGs) with solar metallicity.

First, there is the lack of higher-mass SN progenitors, which suggests that higher mass stars may end their lives as failed SNe \citep{Kochanek08}.  \citet{Smartt09b} and \cite{Smartt15} more clearly demonstrated that the known progenitors of Type IIP SNe have an upper mass limit of $\lesssim18~M_{\odot}$ --- well below the expected mass range for RSG at death.  Although there are alternative hypotheses for the missing SN progenitors, such as post-RSG evolution occuring at lower masses \citep[e.g.,][]{Smith11b,Groh13} or enhanced dust formation prior to core collapse \citep[e.g.,][]{Walmswell12,Beasor16}, the dearth of higher-mass SN progenitors is supported by analyses of stellar populations near SN remnants \citep{Jennings14} and the absence of any Type IIP SNe with the nucleosynthetic signatures of higher-mass ($>20~M_{\odot}$) progenitors \citep{Jerkstrand14}.
Interestingly, the mass range for failed SNe suggested by the missing progenitors corresponds to stars with progenitor structures that make them more difficult to explode \citep{OConnor11,Ugliano12,Pejcha15,Ertl16,Sukhbold16}.

Second, having a significant fraction of core-collapses resulting in failed SNe naturally explains the compact remnant mass function \citep{Kochanek14b,Kochanek15}.
When black holes are formed by ``fall back" onto the proto-neutron star during successful SNe, the mass distributions of black holes and neutron stars are continuous because there is no natural mass scale for the amount of fall back \citep{Zhang08,Fryer12}.  Such continous distributions are inconsistent with observations showing a significant gap between neutron star and black hole masses \citep{Ozel12,Kreidberg12}.  If the core collapse fails to explode a RSG, the resulting black hole mass is the mass of the progenitor's helium core \citep{Lovegrove13}, naturally producing the observed gap between neutron star and black hole mass distributions and the observed black hole masses.

Third, there is evidence that the massive star formation rate may exceed the SN rate (\citealp{Horiuchi11}, but see \citealp{Botticella12} and \citealp{Xiao15}).  
Finally, the recent detection of gravity waves from a pair of merging black holes with masses of $36^{+5}_{-4}~M_{\odot}$ and $29^{+4}_{-4}~M_{\odot}$ \citep{Abbott16a} likely requires the existence of failed SNe \citep{Abbott16b,Belczynski16,Woosley16}.
These lines of evidence for the existence of failed supernovae, while indirect, motivate the direct search for failed SNe.

The formation of a black hole has never been observed and little is known about the range of possible observational signatures.  Some stars likely collapse to form black holes without significant transients \citep{Woosley12}.  However, a failed SN in a RSG likely leads to a visible transient.  \citet{Nadezhin80} suggested, and \citet{Lovegrove13} confirmed with hydrodynamic simulations, that the nearly instantaneous loss of gravitational mass through neutrino emission when a core collapses will lead to a hydrodynamic shock capable of unbinding the loosely bound hydrogen envelope of a RSG.  The resulting optical signature is a shock breakout with $L\sim10^{7}~L_{\odot}$ that lasts for 3-10 days \citep{Piro13} followed by a cool ($\sim3000$ K), $\sim$1 year-long, $\sim$$10^{6}~L_{\odot}$ transient powered by the recombination of the unbound envelope \citep{Lovegrove13}.  
Once sufficiently cool, the slowly expanding ejecta is an ideal environment for dust formation, but this would only occur after the transient has already begun to fade \citep{Kochanek14c}.  Regardless of the nature of any intervening transient the end result is the disappearance of the progenitor.

\citet{Kochanek08} proposed a novel survey to monitor the evolved stars in nearby galaxies to search for failed SNe as disappearing stars.  \citet{Gerke15} presented the results of the first four years of such a survey undertaken with the Large Binocular Telescope (LBT) and found one good failed SN candidate.  This source, in NGC 6946 at RA 20:35:27.56 and Dec +60:08:08.29, which we will hereafter refer to as N6946-BH1, experienced an outburst in 2009 March, first brightening to $\gtrsim$$10^{6}~L_{\odot}$ but then fading to $\sim$$10^{5}~L_{\odot}$ \emph{below} its pre-outburst luminosity.  \citet{Gerke15} found that a coincident source experienced a similar, but slower outburst in the mid-IR.  \citet{Gerke15} also identified the progenitor in earlier archival \emph{Hubble Space Telescope} (\emph{HST}) images with $23.09\pm0.01$ in $F606W$ and $20.77\pm0.01$ in $F814W$.  \citet{Reynolds15} performed a similar search for failed SNe using archival \emph{HST} data and also identified a candidate.  

A new kind of search is vulnerable to new kinds of false positives.  The initial candidate selection was based on a decline in multiple optical bands, but a surviving star could be hidden by dust.
There are several classes of sources known to have transient, heavily-obscured phases.   
Some variable stars, such as R Cor Bor stars, may become optically faint for hundreds of days due to dust forming in their atmospheres \citep{Okeefe39}.  
Stellar mergers can cause the envelope of the primary star to be ejected at low velocities --- ideal conditions for dust formation --- and result in a merger remnant that is luminous in the IR but optically obscured \citep{Crause03,Pejcha16,Pejcha16b}.  
Luminous blue variables may experience eruptive mass loss that obscures a surviving star following a weak transient
--- a SN ``impostor" --- as $\eta$ Carinae did in the mid-1800s \citep[e.g.,][]{Humphreys94,Smith11}.
There are also SN ``impostors" that arise from self-obscured super-AGB stars --- SN 2008S-like transients --- that quickly become re-enshrouded in dust \citep{Prieto08,Kochanek11b,Thompson09}, though some SN ``impostors," from both super-AGB stars and more massive stars, may be lower-luminosity SNe \citep[see][]{Adams15,Adams16}.
Thus, multi-wavelength follow-up is needed to vet failed SN candidates and determine whether the star survived.

In this work we present follow-up observations and analysis of N6946-BH1.  New \emph{HST} imaging confirms that the identified progenitor has disappeared in the optical but a fainter, coincident source is detected in the near-IR (see Fig. \ref{fig:imaging}).  We present the data in \S\ref{sec:data} and describe our spectral energy distribution (SED) modeling in \S\ref{sec:models}.  In \S\ref{sec:results} we present detailed analysis of the progenitor, the outburst, and the post-outburst observational constraints.  While we find the failed SN interpretation for this event to be the most compelling, we consider alternative explanations in \S\ref{sec:alternatives} before closing with our summary and conclusions in \S\ref{sec:summary}.
Following \citet{Gerke15}, we adopt a distance of 5.96 Mpc to NGC 6946 \citep{Karachentsev00} and a Galactic foreground extinction of $E$($B$--$V)=0.303$ based on the \citet{Schlafly11} recalibration of \citet{Schlegel98}.  In this paper, all magnitudes are in the Vega system. 

\section{Data}
\label{sec:data}
\begin{figure*}
                        \includegraphics[width=0.99\textwidth]{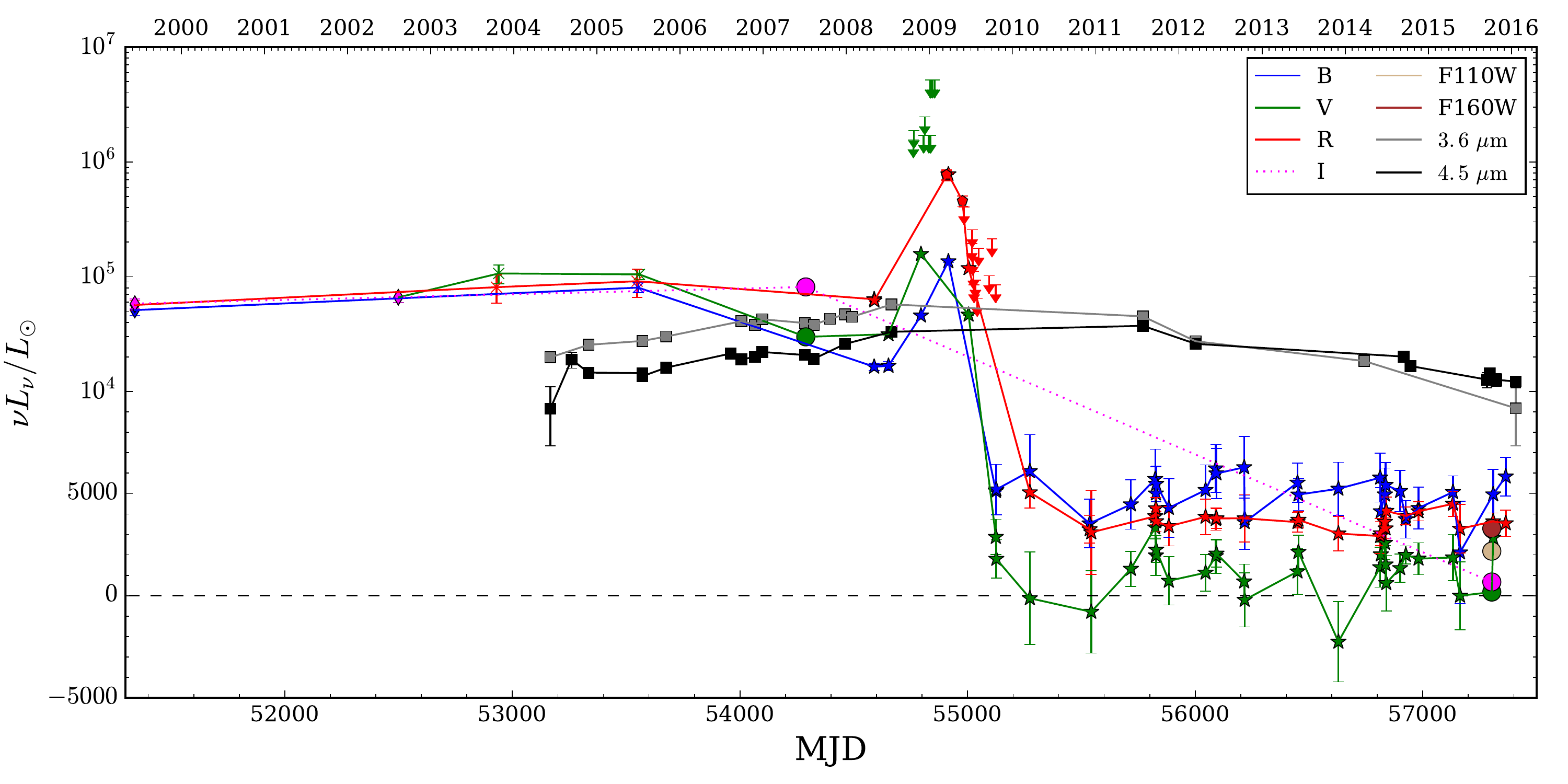}
                      \caption{N6946-BH1 light curves from \emph{HST} (large circles), \emph{SST} (squares), LBT (stars), CFHT (x's), INT (diamonds), PTF (red pentagons and upper limits), and amateur astronomer Ron Arbour (green upper limits).  The vertical axis switches from a linear scale below $10^{4}~L_{\odot}$ to a logarithmic scale above $10^{4}~L_{\odot}$.  A luminosity of zero is given by the dashed black line for comparison.  The uncertainties for the differential LBT and \emph{SST} photometry do not include the significant uncertainty in their ``zeropoints" created by crowding.  The LBT fluxes at late times could easily be zero.  For this purpose, the high resolution \emph{HST} constraints that any remaining optical flux is $<10^{3}~L_{\odot}$ are more relevant.}
                      \label{fig:lc}
                      \end{figure*}

N6946-BH1 was discovered as part of our survey for failed SNe with the LBT \citep{Kochanek08,Gerke15}.  The survey is still ongoing, and in this paper we utilize 38 UBVR epochs from the survey spanning from 3 May 2008 to 8 December 2016 \citep{Adams16s}.  
We calibrated the LBT U-band photometry using \citet{Botticella09}, the B-band using SINGS ancillary data \citep{Kennicutt03}, and the V and R bands from \citet{Welch07}.
We performed image subtraction using {\sc isis} \citep{Alard98,Alard00} with reference images generated by combining the epochs in the top quintile of seeing.  The R and V magnitude ``zeropoints" are based on PSF photometry while B and U come from aperture photometry.  All photometry is presented in Table \ref{tab:photometry}.
We also use image subtraction to place constraints on the late-time variability of N6946-BH1.  
These constraints include an estimate of the systematic uncertainties based on the standard deviation of aperture fluxes measured on a grid of points within $10\arcsec$ of the progenitor location on the difference images after $3\sigma$ clipping for each epoch.
Since the brief optical spike in 2009 the optical fluxes have been consistent with no variability at the level of $\sim$$200~L_{\odot}\>\mathrm{yr}^{-1}$, or roughly $10^{-3}$ of the progenitor flux (see Table \ref{tab:opticalvar} and Fig. \ref{fig:lc}).
\begin{table}
\caption{Late-time Variability Constraints}
\begin{tabular}{cccc}
\hline
\hline
{Filter} & {Variability} & {Date Range} & {Number} \\
 & {[$L_{\odot}\>\mathrm{yr}^{-1}$]} & & {of Epochs} \\
\hline
U              & $-140\pm450$   & 2011-06-04 -- 2015-12-07      &       20      \\
B              & $\hphantom{.} -80\pm160$   & 2011-06-04 -- 2015-05-20      &       24      \\
V              & $\hphantom{.} -10\pm110$   & 2011-06-04 -- 2015-10-13      &       24      \\
R              & $\hphantom{.} -30\pm80\hphantom{1} $     & 2011-09-19 -- 2015-12-07      &       22      \\
\hline
\hline
\label{tab:opticalvar}
\end{tabular}
\end{table}


Since the progenitor was only present in the LBT survey data for three epochs spanning less than 7 months prior to the start of its outburst, we also searched for archival data beyond the \emph{HST} images identified in \citet{Gerke15}.  We found the progenitor in two archival CFHT MegaPipe images \citep{Gwyn08}.  The transformed (to Johnson using the prescription in \citealt{Gwyn08}) MegaCAM apparent aperture magnitudes were $R=21.3\pm0.3$, $V=21.7\pm0.1$, $B=22.8\pm0.1$, and $U=23.5\pm0.2$ on 2 July 2005 and $R=21.5\pm0.3$ and $V=21.7\pm0.2$ on 23 October 2003. 
	There are also archival images of the progenitor from the Isaac Newton 2.5-m Telescope (INT).  For these data we performed PSF photometry using {\sc daophot}.  Calibrating the $r$ and $i$ band data with the Pan-STARRS PS1 catalogs \citep{Chambers16} and the $B$ and $V$ band data by bootstrapping from LBT $B$ and $V$ band photometry, we measure $B = 23.3\pm0.1$, $r = 22.0\pm0.1$, and $i = 21.4\pm0.1$ on 10/12 June 1999 and $V=22.3\pm0.1$ and $i=21.6\pm0.1$ on 12 August 2002.

	We supplement our coverage of the transient with public archival data from the Palomar Transient Factory \citep[PTF;][]{Law09,Ofek12,Laher14} DR2.  We adopt the adaptive aperture ``MAGAUTO" fluxes and upper limits from the PTF catalogs.  The transient is detected in the first available epoch on 17 March 2009 at $R=19.0\pm0.1$ and is last detected at $R=19.6\pm0.1$ on 26 May 2009.
	We also use unfiltered images taken by amateur astronomer Ron Arbour between 5 October 2015 and 23 January 2016 to help constrain the peak magnitude and the start date of the optical transient.  In these images, no source is detected at the position of N6946-BH1 to limiting magnitudes of 17.5 to 18.8.

\begin{figure}
		      \ifpdflatex
                      \includegraphics[width=8.6cm, angle=0]{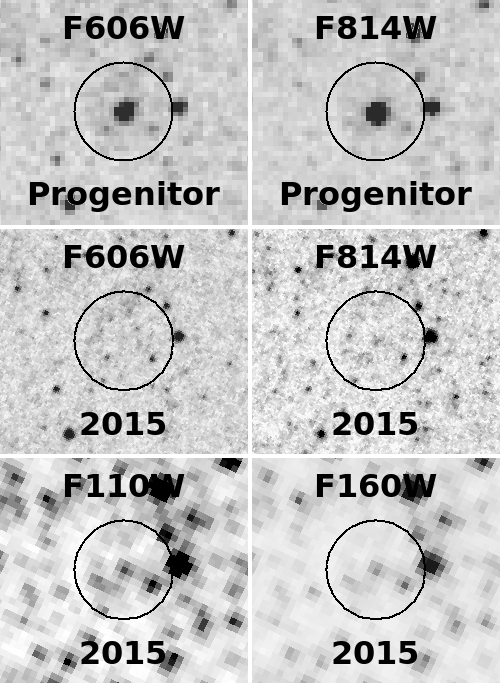}
		      \fi
		      \caption{\emph{HST} images of the region surrounding N6946-BH1.  The top row shows the WFPC2 F606W (left) and F814 (right) progenitor images.  The middle row shows the corresponding 2015 WFC3 images and the bottom row shows WFC3/IR F110W (left) and F160W (right) images.  The circles have a radius of $1\arcsec$.  The progenitor has dramatically faded in the optical but there is still faint near-IR emission. \label{fig:imaging}}
		      \end{figure}

	We utilize both new and archival \emph{HST} data.  For our program, we obtained new Wide Field Camera 3 (WFC3) UVIS $F606W$ and $F814W$ and IR $F110W$ and $F160W$ images on 2015 October 8.  We also use the archival \emph{HST} WFPC2 $F606W$ and $F814W$ images taken 2007 July 8 (PI: M. Meixner, GO-11229) that were considered in \citet{Gerke15}.
	We aligned the new and archival \emph{HST} images using the TWEAKREG and TWEAKBACK tasks in the {\sc drizzlepac} package with RMS errors in the astrometry of $0\farcs04$ for the UVIS images and $0\farcs07$ for the IR images.  We calculated PSF magnitudes for the new \emph{HST} data using the software package {\sc dolphot 2.0} \citep{Dolphin00}\footnote{http://americano.dolphinsim.com/dolphot/} with the same parameter settings as in \citet{Adams15}.  We used the drizzled \emph{HST} WFPC2 $F814W$ pre-outburst image from 2007 as the reference in order to obtain PSF photometry at the progenitor location.  The progenitor has clearly disappeared in the optical (see Fig. \ref{fig:imaging}).  The closest {\sc dolphot} source is $0\farcs045$ from the progenitor position, which is roughly consistent given the astrometric uncertainties.  We also calculated a local World Coordinate System (WCS) alignment between the new and archival \emph{HST} images using the {\sc IRAF} task GEOMAP with RMS errors in the astrometry of $0\farcs012$ for the UVIS images and $0\farcs019$ for the IR images.  Using the new \emph{HST} $F814W$ ($F110W$) as the reference for the local WCS alignment yielded similar photometry for the closest {\sc dolphot} source but now only $0\farcs022$ ($0\farcs017$) from the progenitor location.
	We estimate the likelihood that the \emph{HST} source detected after the outburst is an incidental detection of an unrelated source based on the surface density of similarly bright sources.  The surface density of all {\sc dolphot} sources within $4\arcsec$ of N6946-BH1 is 5.1/arcsec$^{2}$, which corresponds to a 15 percent chance of an unrelated source being detected within $0\farcs045$ (the distance of the closest late-time \emph{HST} source from the progenitor location).  For sources as bright as the detection (in F814W), this drops to a surface density of 1.4/arcsec$^{2}$ and a 0.9\% likelihood.

	We also use both new and archival \emph{Spitzer Space Telescope (SST)} data.  For our program, we obtained new 3.6 and 4.5$~\mu\mathrm{m}$ images taken on 2016 January 21.  We supplemented the IR light curve with avaliable archival images (program IDs: 159, 3248, 10136, 11063, 20320, 20256, 30292, 30494, 40010, 40619, 70040, 80015, 80196, 90124, 10081, 11084, 12000; PIs: J. Andrews, M. Kasliwal, R. Kennicutt, C. Kochanek, R. Kotak, W.P. Meikle, M. Meixner, B. Sugerman).  We performed aperture photometry on the \emph{SST} images using a $2\farcs4$ aperture with a $2\farcs4$--$4\farcs8$ radius sky annulus and the standard aperture corrections from the IRAC instrument handbook.
	Since the \emph{SST} was not designed to have the resolution for extragalactic stellar photometry, it is challenging to accurately measure the absolute photometry of N6946-BH1.  We use {\sc isis} to obtain an accurate measurement of the differential light curve, again estimating systematic uncertainties for each subtracted image from the standard deviation of fluxes on a grid of points within 15$\arcsec$ of N6946-BH1 after $3\sigma$ clipping.
The light curve shows that the IR flux is rising from the first \emph{SST} observations in 2004.  Given the low surface density of variable sources, we attribute the change in IR flux coincident with N6946-BH1 to the source.  The IR flux is significantly above its minimum value at all epochs that we model except for the first (2005-07-02) and last (2016-01-21).  Thus it reasonable to treat the IR flux measurements as detections at every epoch except for those two.
	We estimate the odds that the remaining IR flux in 2016-01-21 could be the result of confusion.  Using a grid of apertures within a 15$\arcsec$ radius of N6946-BH1 we find that $7.7\%$ and $1.8\%$ of the apertures are brighter than our $3.6\mu\mathrm{m}$ and $4.5\mu\mathrm{m}$ measurements, respectively.




		      In the latest epoch (2015 October 8 for \emph{HST} and 2016 Jan 21 for \emph{SST}), it is plausible that only the near-IR photometry are detections of N6946-BH1.  In addition to the IR flux being at or similar to their lowest measurements, {\sc dolphot} reports the `sharpness' parameter for the $F814W$ emission in 2015 to be $-0.48$, which suggests that the emission in this filter is not resolved, and the $S/N$ of the $F606W$ emission is only 2.4.  Thus, we will consider both cases where all photometry in the latest epoch are detections and when all photometry except the near-IR measurements are taken as only upper limits.

		      \section{SED Modeling}
		      \label{sec:models}
		      We model the SED to constrain the physical properties (i.e., luminosity, temperature, mass, and recent mass loss) of the progenitor, its outburst, and the late-time source.  We will use the results of these models to discuss whether the progenitor survived the outburst and whether a failed supernova or some other phenomenon (e.g., stellar merger or eruptive mass loss) best explains the data.

		      Following the methods in \citet{Adams15} and \citet{Adams16}, we model the SED of the progenitor, its outburst, and the late-time source using the dust radiative transfer code {\sc dusty} \citep{Ivezic97,Ivezic99,Elitzur01}.  We use stellar models from \citet{Castelli04} for stars with solar metallicity and effective temperatures between 3500 and 50000 K and revert to blackbody models when attempting to fit temperatures below 3500 K.  We employ a Markov Chain Monte Carlo (MCMC) wrapper around {\sc dusty} to find best-fit models and allowed parameter ranges.  We adopt minimum photometric uncertainties of $10\%$ (to account for uncertainty in distance and metallicity and any systematic problems in the models).  We use silicate dust from \citet{Draine84} with a standard MRN grain size distribution ($\mathrm{d}n/\mathrm{d}a \propto a^{-3.5}$ with 0.005 $\mu\mathrm{m} < a < 0.25~\mu\mathrm{m}$; \citealt{Mathis77}).

		      The IR variability of the progenitor as well as the post-outburst IR emission could be indicative of dust formation.  We consider two modes of mass loss: the ejection of a dusty shell and a steady-state wind.  
		      We assume that all dust formation occurs in the outflowing material once it cools to the dust formation temperature, $T_{\mathrm{f}}\simeq1500$ K.
		      In the shell model, as the shell continues to expand beyond the dust formation radius, $R_{\mathrm{f}}$, the optical depth, $\tau$, decreases, asymptoting at late times to $\tau \propto t^{-2}$, where $t$ is the elapsed time since the ejection of the shell.  For a thin shell, the mass of the ejecta, $M_{\mathrm{ej}}$, corresponding to a given optical depth is given by
		      \begin{equation}
		      M_{\mathrm{ej}}= \frac{4\pi v_{\mathrm{ej}}^{2} t^{2} \tau_{V,\mathrm{tot}}(t)}{\kappa_{V}} ,
		      \label{eqn:mej}
		      \end{equation}
		      where $v_{\mathrm{ej}}$ is the velocity of the ejected shell and $\kappa_{V}$ is the opacity of the dust at $V$ band.  As noted in Tables \ref{tab:progmodels}, \ref{tab:outburstmodels}, \& \ref{tab:latetimemodels}, 
for the shell models we generally fix the ratio between the inner and outer edges of the dust shell, $R_{\mathrm{out}}/R_{\mathrm{in}}$, to 2.  The models where we allow $R_{\mathrm{out}}/R_{\mathrm{in}}$ to vary show that the shell thickness is relatively unconstrained by the data and has little effect on estimates of the other model properties.

		      For a set of post-outburst shell models (labeled as ``with dL/dt" in Table \ref{tab:latetimemodels}) we also include constraints on the late-time variability of the source.  As discussed in \citet{Adams15}, the luminosity of a surviving source of constant intrinsic luminosity is constrained by the variability, $dL_{f,\mathrm{obs}}/dt$, and optical depth of the source in that filter, $f$, according to
		      \begin{equation}
		      L_{*,f}\simeq \frac{1}{2} \frac{t}{\tau_{f,\mathrm{eff}}} \left( \frac{dL_{f,\mathrm{obs}}}{dt} \right) \mathrm{e}^{\tau_{f,\mathrm{eff}}} .
		      \end{equation}
We impose the variability constraints, $dL_{\mathrm{obs}}/dt$, from Table \ref{tab:opticalvar} on the models by adding contributions of 
\begin{equation}
\chi^{2}_{\mathrm{f}} = \left( \frac{dL_{\mathrm{f,obs}}/dt - dL_{\mathrm{f,mod}}/dt}{\sigma_{dL_{\mathrm{f,obs}}}/dt} \right)^{2} 
\end{equation}
for each constrained filter, $\mathrm{f}$, where the model variability, $dL_{\mathrm{f,mod}}/dt$, is
\begin{equation}
\frac{dL_{\mathrm{f,mod}}}{dt} = \frac{2L_{\mathrm{f}} \tau_{\mathrm{f,eff}}}{t} 
\end{equation} 
\citep[see][]{Adams16}.
		      We also consider a set of models where we compare the evolution of the IR flux to the expansion (and cooling) of the dust shell.  For these models we compute the $\chi^{2}$ of a given MCMC step for the latest photometric constraints.  We infer a shell expansion velocity, $v_{\mathrm{ej}}$, based on the elapsed time, $t$, and inner shell edge, $R_{\mathrm{in}}$, of the model.  We then extrapolate the model back to an earlier post-outburst epoch with $\emph{SST}$ observations using this $v_{\mathrm{ej}}$ to find the appropriate $R_{\mathrm{in}}$ for the earlier epoch, generating a new {\sc dusty} model with the optical depth, $\tau$, expected from a $\tau \propto t^{-2}$ scaling, and include the $\chi^{2}$ for this extrapolated model in the MCMC step.

		      For the wind scenario, the inner edge of the dust is set by the formation radius $R_{\mathrm{f}}$ and we allow the thickness of the dust `shell' to vary.  Since the optical depth of a wind (or shell) is dominated by the inner edge, the results are usually insensitive to the thickness $R_{\mathrm{out}}/R_{\mathrm{in}}$.
		      The mass-loss rate needed to produce a given optical depth is 
		      \begin{equation}
		      \dot{M} = \frac{4\pi v_{\mathrm{w}} R_{\mathrm{f}} \tau_{V,\mathrm{tot}}}{\kappa_{V}} 
		      \label{eqn:mdot}
		      \end{equation}
		      where $v_{\mathrm{w}}$ is the velocity of the wind.

	\section{Results}
	\label{sec:results}
	We will, in turn, consider the implications of the SED constraints for the progenitor, the optical outburst, and the post-outburst fate of the star.
The models for these three phases are summarized in Tables \ref{tab:progmodels}, \ref{tab:outburstmodels}, and \ref{tab:latetimemodels}.

	\subsection{Progenitor}
	As shown in Fig. \ref{fig:lc}, observations over the 10 years prior to the optical outburst reveal a very luminous ($10^{5.3}~L_{\odot}$) star that remained at a constant brightness between 1999 and 2005, but then faded in the optical between 2005 and 2008 as it brightened at $3.6\>\mu\mathrm{m}$.  We fit the SED for three progenitor epochs: summer 2005 with archival CFHT data (2 July 2005) and the nearest corresponding archival \emph{SST} observations (20 July 2005), summer 2007 with archival \emph{HST} (8 July 2007) and the nearest corresponding archival \emph{SST} observation (3 July 7 2007), and summer 2008 with the difference between the LBT pre-outburst images (3-4 May 2008 and 5 July 2008) and post-outburst images ($\sim$25 epochs between (4 June 2011 and 7 December 2015) together with archival SST observations from that summer (7 July 2008)\footnote{We use the difference since the effect of crowding in ground-based photometry is likely larger than the small flux remaining in the late-time \emph{HST} imaging.  However, we note that adopting instead the pre-outburst LBT constraints has little effect on the results.}.  The constraints from these fits are given in Table \ref{tab:progmodels}.  

			\begin{figure}
                        \includegraphics[width=8.6cm, angle=0]{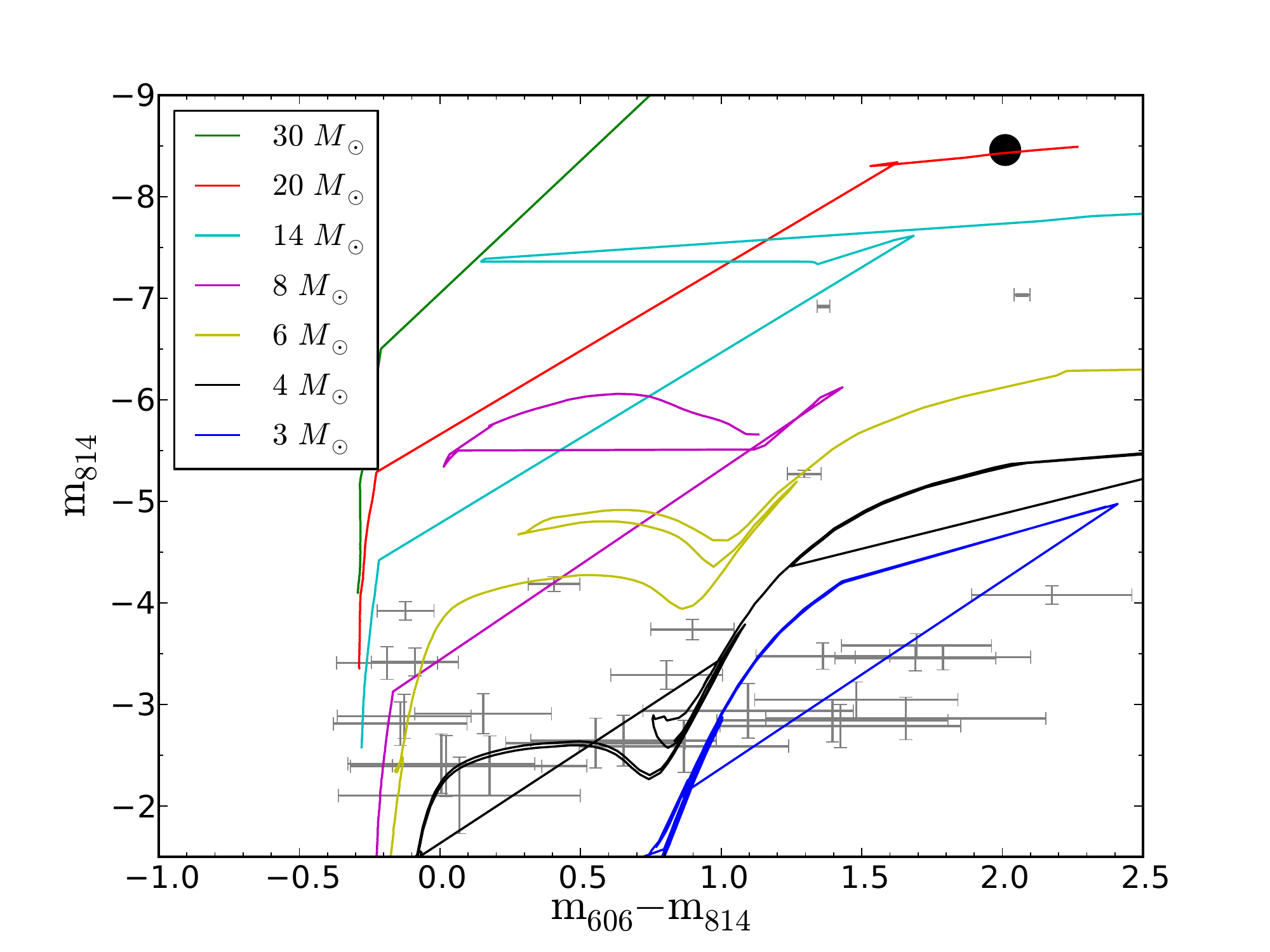}
			\caption{Color (absolute) magnitude diagram of stars within $1\farcs8$ (50 pc projected) of N6946-BH1, corrected for Galactic extinction and assuming no additional extinction in the local environment of NGC 6946, are shown in gray. The progenitor is shown by the large black circle (with the uncertainties smaller than the size of the symbol).  For comparison, evolutionary tracks for solar metallicity stars of various initial masses are shown as solid lines \citep{Bressan12,Chen15}.\label{fig:cmd}}
			\end{figure}

			In Fig. \ref{fig:cmd} we show a color magnitude diagram of the bright stellar sources within a projected radius of 50 pc from N6946-BH1, corrected for Galactic extinction, along with evolutionary tracks for the {\sc parsec} (v. 1.2S) stellar models \citep{Bressan12,Chen15} of solar metallicity stars of various initial masses.  The location of the theoretical main-sequence relative to the location of the nearby stars in the color-magnitude diagram constrains the extinction in the local environment of N6946-BH1 to be $E$($B$--$V)\lesssim0.2$ mag.  This is consistent with the constraints from the MCMC SED modeling of the progenitor which give $E$($B$--$V)\lesssim0.15$ (see models P1-3 in Table \ref{tab:progmodels}).  Accordingly we adopt $E$($B$--$V)=0$ for the local environment in all other models (unless noted otherwise).

			\begin{figure}
                        \includegraphics[width=8.6cm, angle=0]{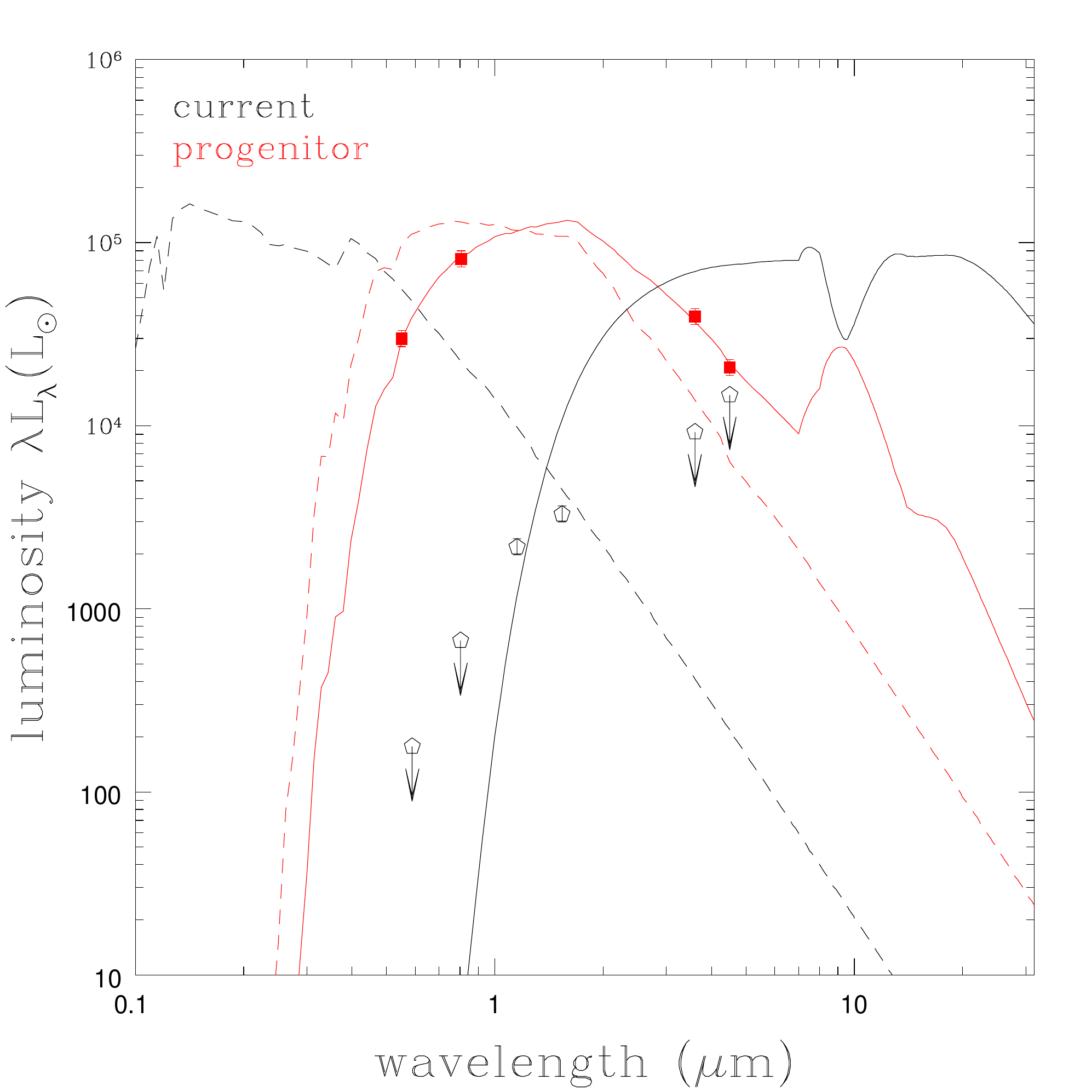}
			\caption{The SED of the progenitor of N6946-BH1 in 2007 (red) and of a possible surviving star in 2016 obscured by a dusty wind (black).  The solid red line gives the best-fitting model (from P5 in Table \ref{tab:progmodels}) for the progenitor (log $L_{*}/L_{\odot}=5.29$, $T_{*}=4480$ K, $\tau_{V,\mathrm{tot}}=3.3$, $T_{\mathrm{d}}=1800$ K, and $\chi^{2}=0.8$) and the dashed red line gives the corresponding unobscured spectrum.  The red squares give the \emph{HST} and \emph{SST} constraints in 2007.  The open black pentagons give the latest \emph{HST} and \emph{SST} constraints on a surviving star when treating our $F110W$ and $F160W$ photometry as detections and the other bands as upper limits.  The best-fitting model (from L25 in Table \ref{tab:latetimemodels} for a surviving star with the luminosity of the progenitor (log $L_{*}/L_{\odot}=5.3$) obscured by a dusty wind ($T_{\mathrm{f}}=1500$ K) and shown by the solid black line has $T_{*}=14500$, $\tau_{V,\mathrm{tot}}=37.3$, $R_{\mathrm{out}}/R_{\mathrm{in}}=72$ and a $\chi^{2}$ of 264.  Clearly, this `best-fit' model is in gross disagreement with the latest photometric constraints.  A surviving star cannot be hidden by a dusty wind because the hot dust that dominates the obscuration reradiates the stellar emission in the near to mid-IR.  In order to hide the luminosity of the progenitor, the bulk of the emission must be radiated by cooler dust at wavelengths redward of $4.5~\mu\mathrm{m}$.
			\label{fig:sed_wind}}
			\end{figure}

	Good fits to the progenitor data are only obtained when allowing for the presence of circumstellar dust (models P1-6).  We show the best-fit progenitor model for summer 2007 (model P5) in Fig. \ref{fig:sed_wind}.  The low stellar temperatures and hot dust temperatures in the three progenitor epochs are consistent with a RSG emitting a dusty wind.  The three epochs of progenitor fits hint that the progenitor may have had a decreasing effective temperature and a slightly increasing mass loss rate, but given the uncertainties it is difficult to make a definitive statement.  Between the 2007 and 2008 epochs, the IR luminosity increased more than the optical luminosity decreased, suggesting that the bolometric luminosity may have started to rise months before the observed optical transient.

	\begin{figure}
        \includegraphics[width=8.6cm, angle=0]{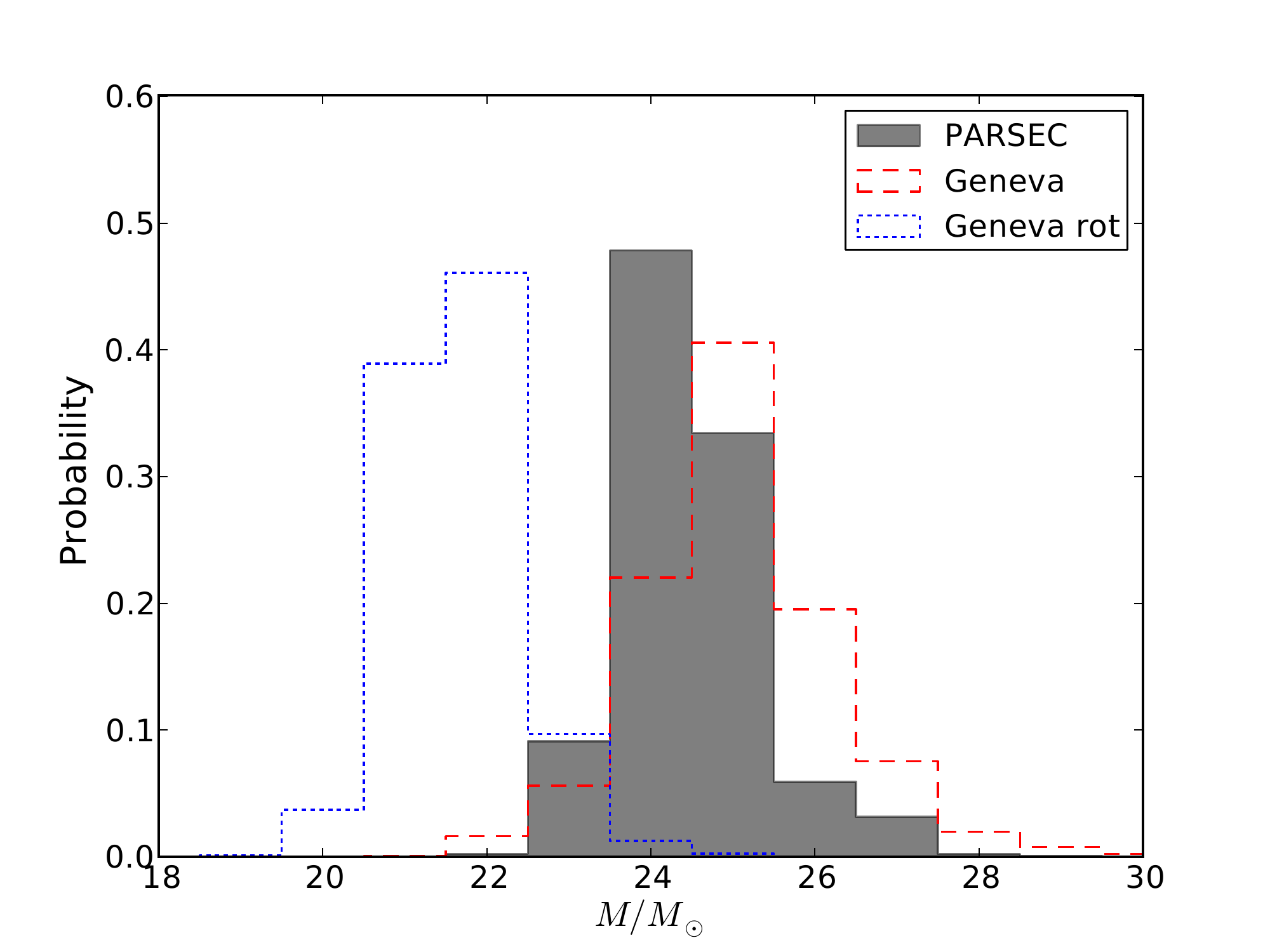}
	\caption{Progenitor mass probability distribution function based on MCMC fits to the pre-outburst \emph{HST} and \emph{SST} photometry (model P2) and the PARSEC (shaded gray) and Geneva rotating (blue dotted line) and non-rotating (red dashed line) stellar models when allowing $E(B-V)$ and $R_{\mathrm{out}}/R_{\mathrm{in}}$ to vary. \label{fig:mass_pdf}}
	\label{fig:progmass}
	\end{figure}

	We matched the estimated progenitor luminosity and effective temperature estimates from the MCMC realizations with those from the {\sc parsec} and Geneva stellar models \citep{Ekstrom12,Yusof13}.  The {\sc parsec} and the non-rotating Geneva models yield a probability distribution for the progenitor mass peaking at $\sim25~M_{\odot}$ while the distribution for the rotating Geneva models peaks at a slightly lower mass of $\sim22~M_{\odot}$ (see Fig. \ref{fig:mass_pdf}).
This mass lies directly within the regime expected by the missing RSG problem \citep{Kochanek08,Smartt09b,Smartt15} and corresponds to a progenitor with a high ``compactness" parameter that is considered most likely to result in a failed supernova \citep{OConnor11,Ugliano12,Horiuchi14,Sukhbold14,Pejcha15,Sukhbold16}.

        \begin{table*}
\begin{center}
\begin{minipage}{16cm}
\caption{Progenitor SED Models}
\label{tab:progmodels}
\begin{tabular}{llllllllll}
\hline
\hline
{Model} & {Date} & {log $L/L_{\odot}$} & {$T_{*}$ [K]} & {$\tau_{V,\mathrm{tot}}$} & {$T_{\mathrm{d}}$ [K]} & {$R_{\mathrm{out}}/R_{\mathrm{in}}$} & {log $M_{\mathrm{ej}}/M_{\odot}$} & {$E$($B$--$V$)} & {$\chi^{2}_{\mathrm{min}}$} \\
\hline
P1 &
2005-07-02 &
$5.31^{+0.12}_{-0.07}$ &	
$6700^{+1200}_{-990}$ &	
$1.4^{+0.4}_{-0.4}$ &	
$1720^{+240}_{-400}$ &	
$2.5^{+26.3}_{-1.3}$ &	
$-6.8^{+0.2}_{-0.1}$ &	
$0.01^{+0.14}_{-0.01}$ &	
$1.5$ 	\\
P2 &
2007-07-08 &
$5.30^{+0.05}_{-0.05}$ &	
$4310^{+750}_{-380}$ &	
$3.4^{+0.9}_{-1.0}$ &	
$1580^{+360}_{-520}$ &	
$7.9^{+71.5}_{-6.5}$ &	
$-6.5^{+0.3}_{-0.2}$ &	
$0.00^{+0.05}_{-0.00}$ &	
$0.5$ 	\\

P3 &
2008-07-05 &
$5.35^{+0.36}_{-0.07}$ &	
$5580^{+850}_{-2060}$ &	
$3.9^{+1.3}_{-1.9}$ &	
$1550^{+390}_{-1490}$ &	
$2.6^{+67.0}_{-1.3}$ &	
$-6.2^{+2.8}_{-0.2}$ &	
$0.01^{+0.02}_{-0.01}$ &	
$0.5$ 	\\

P4 &
2005-07-02 &
$5.28^{+0.07}_{-0.06}$ &	
$6610^{+1150}_{-940}$ &	
$1.4^{+0.2}_{-0.2}$ &	
$1740^{+220}_{-400}$ &	
2 (fixed) &	
$-4.6^{+0.4}_{-0.2}$ &	
0 (fixed) &	
$2.1$ 	\\
P5 &
2007-07-08 &
$5.29^{+0.04}_{-0.06}$ &	
$3260^{+1670}_{-320}$ &	
$1.5^{+2.0}_{-1.4}$ &	
$1300^{+590}_{-1240}$ &	
2 (fixed) &	
$-4.3^{+4.9}_{-1.3}$ &	
0 (fixed) &	
$0.8$ 	\\
P6 &
2008-07-05 &
$5.55^{+0.12}_{-0.06}$ &	
$3590^{+160}_{-80}$ &	
$2.0^{+0.5}_{-0.4}$ &	
$940^{+800}_{-530}$ &	
2 (fixed) &	
$-3.6^{+1.4}_{-1.1}$ &	
0 (fixed) &	
$0.5$ 	\\
P7 &
2007-07-08 &
$5.31^{+1.5}_{-0.01}$ &	
$3010^{+6310}_{-120}$ &	
0 (fixed) &	
... &	
... &	
... &	
$0.02^{+1.74}_{-0.01}$ &	
$4.7$ 	\\
P8 &
2005-07-02 &
$5.53^{+0.01}_{-0.01}$ &	
$4350^{+100}_{-110}$ &	
0 (fixed) &	
... &	
... &	
... &	
0 (fixed) &	
$33.2$ 	\\
P9 &
2007-07-08 &
$5.30^{+0.01}_{-0.01}$ &	
$2960^{+90}_{-90}$ &	
0 (fixed) &	
... &	
... &	
... &	
0 (fixed) &	
$5.4$ 	\\
P10 &
2008-07-05 &
$5.45^{+0.02}_{-0.02}$ &	
$2900^{+60}_{-60}$ &	
0 (fixed) &	
... &	
... &	
... &	
0 (fixed) &	
$15.7$ 	\\
\hline
\hline
\end{tabular}
\end{minipage}
\end{center}
Constraints from MCMC models of the progenitor SED.  The uncertainties give the 90\% confidence intervals.  $L_{*}$ is the bolometric luminosity of the source, $T_{*}$ is the intrinsic effective temperature of the input SED, $T_{\mathrm{d}}$ is the dust temperature at the inner radius of the dust shell ($R_{\mathrm{in}}$), $R_{\mathrm{out}}/R_{\mathrm{in}}$ is the thickness of the dust shell, $E$($B$--$V$) is the local $E$($B$--$V$) in NGC 6946 in addition to $E$($B$--$V)=0.303$ adopted for Galactic extinction, and $\chi^{2}$ is the fit of the model.  For the $R_{\mathrm{out}}/R_{\mathrm{in}}=2$ models $M_{\mathrm{ej}}$ is the ejected mass implied by Eqn. \ref{eqn:mej}, otherwise it is the mass loss rate $M_{\odot}\>\mathrm{yr}^{-1}$ implied by Eqn. \ref{eqn:mdot}.  Local extinction in NGC 6946 is not required to achieve good fits, but circumstellar dust is.
\end{table*}


	\subsection{Outburst}
	\label{sec:outburst}

	The outburst is not well-sampled in time, but we can set useful constraints on the duration and minimum peak luminosity of the event.  The source appeared quiescent in LBT observations on 5 July 2008, but had brightened by a factor of several in observations taken on 25 November 2008. The peak observed (apparent) magnitudes occurred in the following LBT epoch on 25 March 2009 with $R=19.0$, $V=20.0$, and $B=22.1$, corresponding to luminosities ($\nu L_{\nu}$) of $7.8\times 10^5~L_{\odot}$, $4.9\times 10^{5}~L_{\odot}$, and $1.4\times 10^{5}~L_{\odot}$ in $R$, $V$, and $B$ respectively (unfortunately there was no $U$-band data for this epoch).  At the next LBT epoch on 6 June 2009, the transient had faded to $R=21.0$, $V=22.6$, $B=24.4$, and $U=24.7$, and by the subsequent LBT epoch on 20 Oct 2009 the source had already dropped below the progenitor flux in all optical bands (see Fig. \ref{fig:lc}).
	
PTF observations show that the source was present and likely already declining in flux from the earliest available images taken 17 March 2009.  The transient is undetected in images from R. Arbour (private communication) with limiting magnitudes of $\sim18.7$ in early January 2009.
	Thus, the optical transient likely began between late November 2008 and mid-March 2009 and became optically faint between early June 2009 and late October 2009,  constraining the duration of the optical transient to be between 3 and 11 months.
	Unfortunately, there are no \emph{SST} observations during the optical outburst, but it seems clear that the transient evolved more slowly in the IR than in the optical.

	We generated models of the SED for the two LBT epochs taken during the optical outburst (see Table \ref{tab:outburstmodels}).  Without IR constraints there is a significant degeneracy, with hotter intrinsic temperatures yielding larger bolometric luminosities obscured by larger optical depths.  Thus, we also present models with $T_{*}\equiv 3500$ since, in addition to being the likely temperature of the progenitor, the color temperatures for the failed SN transients powered by hydrogen recombination in the models of \citet{Lovegrove13} are relatively cool.
	The peak luminosity was at least $10^{6}~L_{\odot}$ (see model O1 in Table \ref{tab:outburstmodels}).
	The bolometric luminosity returned to roughly that of the progenitor by the first \emph{SST} epoch after the peak of the optical outburst, which, unfortunately, was not taken until 27 July 2011.  

	We can also use the minimum and maximum durations of the optical transient to set constraints on the velocity of material ejected during the event if we assume that the optical flux collapses before the IR flux due to the formation of dust in the newly ejected material.  This seems a reasonable assumption given that all models of the SED after the optical disappearance of the source (except for shell models for 2012-03-16 and 2011-07-28 that allow very hot stellar temperatures) require much higher optical depths than models of the progenitor in order to produce good fits.
	Given the SED modeling constraints on the source luminosity and temperature on 2011-07-28 and $T_{\mathrm{f}}=1500$ K ($L_{*}$, $T_{*}$, and $T_{\mathrm{d}}$ set $R_{\mathrm{f}}$), material ejected at the start of the transient must have $170<v_{\mathrm{ej}}<560~\mathrm{km}\>\mathrm{s}^{-1}$ in order to reach $R_{\mathrm{f}}$ and form dust to extinguish the optical flux within the time window of the observational constraints.
A moderately hotter dust condensation temperature of $T_{\mathrm{f}}=2000$ would decrease the minimum velocity to $v_{\mathrm{ej}}\simeq75~\mathrm{km}\>\mathrm{s}^{-1}$.

	These constraints on the peak luminosity, possible duration, and ejected velocities are broadly consistent with the cool ($\sim3000$ K), $\sim$1 yr, $\sim$$10^{6}~L_{\odot}$ transient with $v_{\mathrm{ej}}\sim100~\mathrm{km}\>\mathrm{s}^{-1}$ predicted by \citet{Lovegrove13} for a failed supernova in a RSG unbinding the hydrogen envelope. 
	Of course, with the very limited observations, the outburst may also be consistent with other types of events, such as a SN impostor or a stellar merger.
	Thus, we must rely on the late-time evolution of the source to decipher its true nature.

        \begin{table*}
\begin{center}
\begin{minipage}{16cm}
\caption{SED Outburst Models}
\label{tab:outburstmodels}
\begin{tabular}{lllllllll}
\hline
\hline
{Model} & {Date} & {log $L/L_{\odot}$} & {$T_{*}$ [K]} & {$\tau_{V,\mathrm{tot}}$} & {$T_{\mathrm{d}}$ [K]} & {$R_{\mathrm{out}}/R_{\mathrm{in}}$} & {log $M_{\mathrm{ej}}/M_{\odot}$} & {$\chi^{2}_{\mathrm{min}}$} \\
\hline
O1 &
2009-03-25 &
$6.33^{+0.13}_{-0.09}$ &	
$3300^{+180}_{-210}$ &	
0 (fixed) &	
... & 
... & 
... & 
$0.9$ 	\\
O2 &
2009-06-22 &
$5.47^{+0.15}_{-0.12}$ &	
$3190^{+230}_{-220}$ &	
0 (fixed) &	
... &	
... &	
... &	
$5.1$ 	\\
O3 &
2009-03-25 &
$6.51^{+2.00}_{-0.26}$ &	
$5080^{+27670}_{-1750}$ &	
$3.2^{+4.3}_{-2.8}$ &	
$280^{+1180}_{-230}$ &	
2 (fixed) &	
$\hphantom{-}0.5^{+5.3}_{-3.9}$ &	
$0.0$ 	\\
O4 &
2009-06-22 &
$6.54^{+1.4}_{-1.11}$ &	
$12010^{+30960}_{-8810}$ &	
$6.5^{+1.5}_{-6.2}$ &	
$390^{+1450}_{-330}$ &	
2 (fixed) &	
$-0.1^{+5.2}_{-4.2}$ &	
$0.0$ 	\\
O5 &
2009-03-25 &
$6.79^{+0.14}_{-0.13}$ &	
3500 (fixed) &	
$2.0^{+0.7}_{-0.7}$ &	
$240^{+1230}_{-190}$ &	
2 (fixed) &	
$-0.3^{+3.7}_{-2.9}$ &	
$2.5$ 	\\
O6 &
2009-06-22 &
$5.94^{+0.14}_{-0.13}$ &	
3500 (fixed) &	
$2.4^{+0.6}_{-0.6}$ &	
$310^{+1240}_{-260}$ &	
2 (fixed) &	
$-1.5^{+4.0}_{-2.5}$ &	
$2.4$ 	\\
\hline
\hline
\end{tabular}
\end{minipage}
\end{center}
Constraints from MCMC models of the SED during the outburst.  The columns are the same as in Table \ref{tab:progmodels}, except the $E(B-V)$ is now fixed to the Galactic extinction for NGC 6946.
\end{table*}

	\subsection{Late-Time}
	\label{sec:latetime}

\begin{figure}
        \includegraphics[width=8.6cm, angle=0]{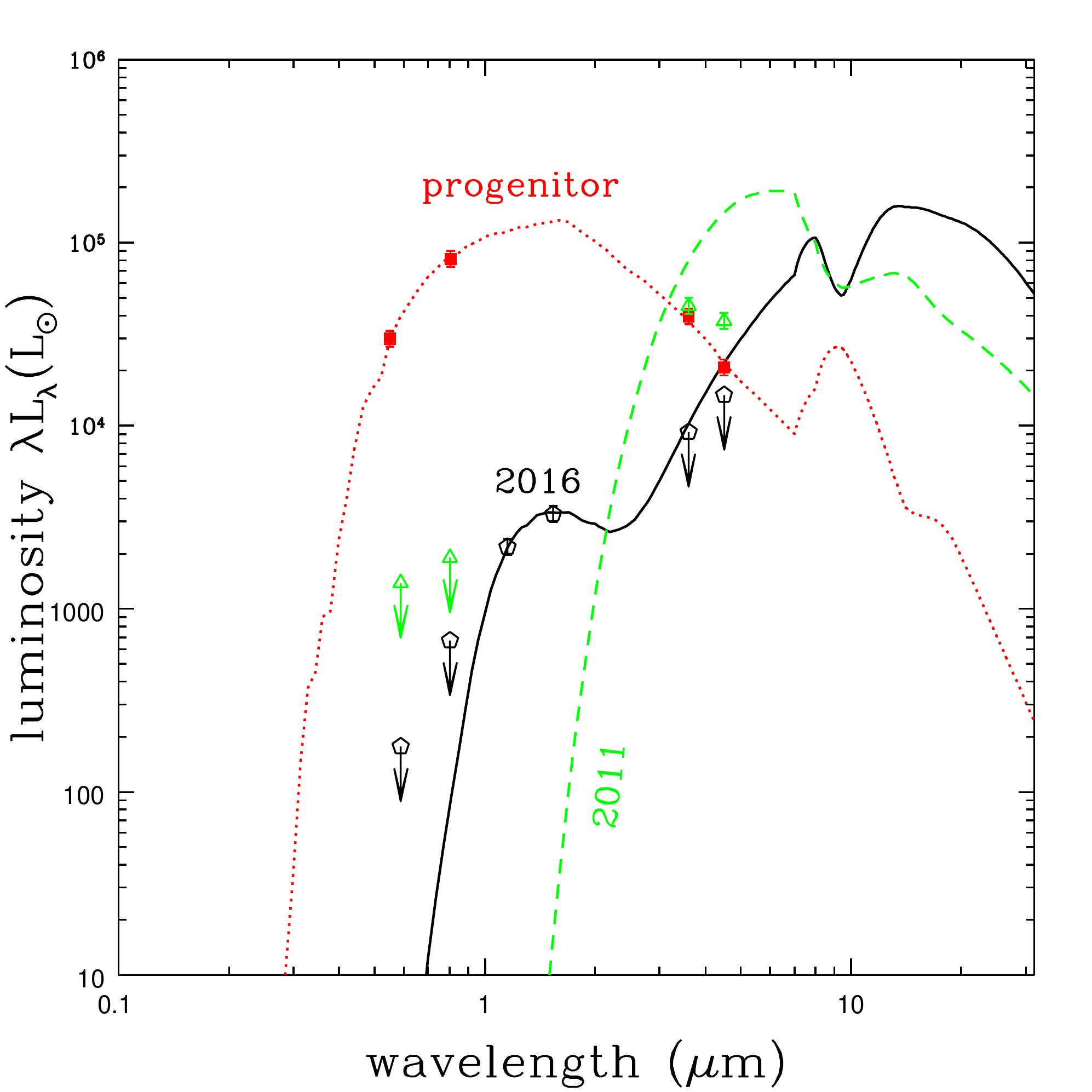}
	\caption{The SED of N6946-BH1 in 2016 with the best-fitting model (from L30 in Table \ref{tab:latetimemodels}) for a surviving star with the luminosity of the progenitor (log $L_{*}/L_{\odot}=5.3$) obscured by a dusty shell when treating our $F110W$ and $F160W$ photometry as detections and the other bands as upper limits is given by the solid black line.  The open black pentagons are the latest SED constraints from \emph{HST} and \emph{SST}.  
		This particular model has $T_{*}=10700$ K, $\tau_{V,\mathrm{tot}}=24.2$, $T_{\mathrm{d}}=570$ K and $\chi^{2}=7.8$.  The dashed green line gives the corresponding shell model for 28 July 2011 based on the implied $v_{\mathrm{e}}$ ($\tau_{V,\mathrm{tot}}=206$ and $T_{\mathrm{d}}=1400$ K) and the green triangles give the \emph{SST} and LBT constraints for that epoch.  The high optical depth required by the 2016 models correspond to such high optical depths in 2011 that the models are optically thick even at $3.6~\mu\mathrm{m}$, leading to a step drop in the SED blueward of $4.5~\mu\mathrm{m}$ that is in gross contradiction with the observational constraints.  For comparison, the progenitor constraints and best-fit model are given by the red squares and dotted line. \label{fig:sed_shell}}
		\end{figure}

	The LBT difference imaging shows that the optical flux collapsed by 20 October 2009 and has remained far below the flux of the progenitor since then (see Fig. \ref{fig:lc}).  This result is confirmed by the late-time \emph{HST} data from 8 October 2015 showing that the progenitor has clearly vanished in the optical, where the emission decreased by at least 5 magnitudes (see Fig. \ref{fig:imaging}, \ref{fig:sed_wind}, and \ref{fig:sed_shell}).  The mid-IR emission has evolved much more slowly than the optical, but the source is now fainter than the progenitor at $3.6\mu\mathrm{m}$ and nearly as faint at $4.5\mu\mathrm{m}$.  There is still faint near-IR emission coincident with the progenitor location (unfortunately there is no progenitor near-IR data).  We will present SED models for three late-time epochs for which there are both $3.6\mu\mathrm{m}$ and $4.5\mu\mathrm{m}$ observations: 28 July 2011, 16 March 2012, and 21 January 2016 (see Table \ref{tab:latetimemodels}).  
We compliment these \emph{SST} constraints with optical $V$ and $R$-band constraints inferred from the \emph{HST} ($F606W$ and $F814W$) photometry on 8 October 2015 extrapolated to these three dates using the LBT constraints on the variability of the $V$ and $R$-band between 2011 and 2015 listed in Table \ref{tab:opticalvar}.

	First, we consider whether the progenitor could have survived as a heavily obscured star with a steady-state wind or an ejected shell, given the late-time photometric constraints.
	Models with the progenitor surviving behind a dusty wind do not fit the data well even if the surviving star's temperature is allowed to vary and 
	all of the photometry is treated as upper limits except for the clear \emph{HST} detections in the near-IR (see model L24 in Table \ref{tab:latetimemodels} and Fig. \ref{fig:sed_wind}).
	The basic issue is that even at high optical depths the models are unable to reradiate the progenitor's luminosity primarily at wavelengths redward of $4.5~\mu\mathrm{m}$, as required by the photometric constraints, because the bulk of the light is reprocessed by hot dust that produces flux in the near-IR (see Fig. \ref{fig:sed_wind}).  The 4.5 $\mu\mathrm{m}$ constraint can only be avoided if the characteristic dust temperature is $\lesssim600$ K, far below typical dust formation temperatures \citep[see, e.g.,][]{Gail14}.  Thus, the progenitor did not survive behind a thick, dusty wind.  

		Models with the progenitor surviving behind an ejected shell are also inconsistent with the observational constraints.  A $T_{*}\sim3500$ K progenitor hidden behind dust cannot easily reproduce the small slope between the near-IR constraints.  A much hotter surviving star ($T_{*}\sim$14000 K) obscured by a dusty shell (model L23) is better able to match the near-IR constraints, although still with a best $\chi^{2}$ of 40 when including the variability constraints.
		Good fits can only be achieved by (in addition to allowing $T_{*}$ to vary) treating all of the photometry as upper limits except for the near-IR.
		In this case, a model with a hot star ($T_{*}\sim11000$ K) and a high optical depth ($\tau\sim24$) from cool dust ($T_{\mathrm{d}}\sim170$ K) can fit the data with $\chi^{2}\simeq0.7$ (see L25 in Table \ref{tab:latetimemodels}).  However, such a cool dust temperature requires a much higher ejecta velocity ($\sim5000~\mathrm{km}\>\mathrm{s}^{-1}$) than allowed by the constraints on the elapsed time between the start of the optical outburst and the collapse of the optical flux due to dust formation ($170<v_{\mathrm{ej}}<560~\mathrm{km}\>\mathrm{s}^{-1}$; see \S\ref{sec:outburst}).  
		Enforcing a prior on the ejecta velocity (with a 10\% uncertainty) increases the dust temperature and also worsens the fit to $\chi^{2}\sim7.8$ (see L30 in Table \ref{tab:latetimemodels} and Fig. \ref{fig:sed_shell}).

		Another consideration is the evolution of the IR flux.  The late-time $3.6\mu\mathrm{m}$ flux decreases faster than the $4.5\mu\mathrm{m}$ flux.  Although this would be a natural consequence of the dust temperature decreasing as the ejected shell expands, the corresponding evolution of the optical depth is problematic for the models.
		The high optical depths required to fit the 2016 photometric constraints  would correspond to high enough optical depths in 2011 to be optically thick even into the mid-IR, resulting in an SED in 2011 that would drop steeply blueward of $4.5~\mu\mathrm{m}$, in gross disagreement with the \emph{SST} photometry from 2011 (see L32 in Table \ref{tab:latetimemodels} and Fig. \ref{fig:sed_shell}).

		The conclusion we draw from this section is that the evolution of the SED can only be well-fit by our models if the bolometric luminosity fades to be significantly below that of the progenitor.  The high optical depth required by the late-time photometric constraints makes the result essentially independent of the stellar temperature or the details of the atmospheric models.
		What then would explain the late-time flux?

        \begin{table*}
\begin{center}
\begin{minipage}{16cm}
\caption{Late-time SED Models}
\label{tab:latetimemodels}
\begin{tabular}{llllllllll}
\hline
\hline
{Model} & {Date} & {log $L/L_{\odot}$} & {$T_{*}$ [K]} & {$\tau_{V,\mathrm{tot}}$} & {$T_{\mathrm{d}}$ [K]} & {$R_{\mathrm{out}}/R_{\mathrm{in}}$} & {log $v_{\mathrm{ej}}$/km s$^{-1}$} & {log $M_{\mathrm{ej}}/M_{\odot}$} & {$\chi^{2}_{\mathrm{min}}$} \\
\hline
L1 &
2011-07-28 &
$5.10^{+0.53}_{-0.14}$ &	
3500 (fixed) &	
$14.1^{+7.1}_{-6.1}$ &	
$1000^{+710}_{-600}$ &	
2 (fixed) &	
$2.1^{+1.0}_{-0.5}$ &	
$-3.1^{+2.2}_{-0.9}$ &	
$1.0$ 	\\
L2 &
2012-03-16 &
$4.99^{+0.39}_{-0.20}$ &	
3500 (fixed) &	
$14.2^{+6.1}_{-5.2}$ &	
$780^{+560}_{-350}$ &	
2 (fixed) &	
$2.2^{+0.7}_{-0.5}$ &	
$-2.8^{+1.4}_{-1.0}$ &	
$1.1$ 	\\
L3 &
2016-01-21 &
$4.24^{+0.02}_{-0.02}$ &	
3500 (fixed) &	
$13.6^{+0.8}_{-0.7}$ &	
$1020^{+80}_{-70}$ &	
2 (fixed) &	
$1.2^{+0.1}_{-0.1}$ &	
$-4.0^{+0.2}_{-0.2}$ &	
$335$ 	\\
L4 &
2016-01-21 &
$4.68^{+0.52}_{-0.18}$ &	
$10390^{+3080}_{-1650}$ &	
$12.8^{+1.9}_{-2.7}$ &	
$850^{+350}_{-260}$ &	
2 (fixed) &	
$2.0^{+0.7}_{-0.3}$ &	
$-2.4^{+1.4}_{-0.7}$ &	
$25.9$ 	\\
L5 &
2011-07-28 &
5.30 (fixed) &	
$2300^{+610}_{-190}$ &	
$12.6^{+6.3}_{-7.2}$ &	
$170^{+440}_{-120}$ &	
2 (fixed) &	
$3.6^{+1.5}_{-1.0}$ &	
$-0.2^{+3.0}_{-2.2}$ &	
$1.0$ 	\\
L6 &
2012-03-16 &
5.30 (fixed) &	
$15870^{+29310}_{-13090}$ &	
$5.6^{+13.9}_{-5.5}$ &	
$720^{+410}_{-430}$ &	
2 (fixed) &	
$2.8^{+0.3}_{-0.2}$ &	
$-1.9^{+0.9}_{-1.7}$ &	
$0.5$ 	\\
L7 &
2016-01-21 &
5.30 (fixed) &	
$14430^{+850}_{-810}$ &	
$14.4^{+1.0}_{-1.0}$ &	
$550^{+20}_{-20}$ &	
2 (fixed) &	
$2.8^{+0.0}_{-0.0}$ &	
$-0.8^{+0.1}_{-0.1}$ &	
$27.3$ 	\\
L8 &
2011-07-28 &
5.30 (fixed) &	
3500 (fixed) &	
$9.8^{+6.8}_{-4.6}$ &	
$710^{+420}_{-110}$ &	
2 (fixed) &	
$2.5^{+0.2}_{-0.4}$ &	
$-2.5^{+0.5}_{-1.1}$ &	
$4.5$ 	\\
L9 &
2012-03-16 &
5.30 (fixed) &	
3500 (fixed) &	
$15.4^{+5.5}_{-5.8}$ &	
$490^{+60}_{-40}$ &	
2 (fixed) &	
$2.7^{+0.1}_{-0.1}$ &	
$-1.7^{+0.3}_{-0.4}$ &	
$1.7$ 	\\
L10 &
2016-01-21 &
5.30 (fixed) &	
3500 (fixed) &	
$24.6^{+0.7}_{-0.7}$ &	
$340^{+30}_{-40}$ &	
2 (fixed) &	
$2.7^{+0.1}_{-0.1}$ &	
$-0.8^{+0.2}_{-0.1}$ &	
$1480$ 	\\
L11 &
2011-07-28 &
5.30 (fixed) &	
3500 (fixed) &	
$7.4^{+2.3}_{-2.1}$ &	
1500 (fixed) &	
$21.8^{+57.7}_{-19.0}$ &	
$1.8^{+0.0}_{-0.0}$ &	
$-6.2^{+0.1}_{-0.1}$ &	
$10.3$ 	\\
L12 &
2012-03-16 &
5.30 (fixed) &	
3500 (fixed) &	
$3.0^{+1.2}_{-1.0}$ &	
1500 (fixed) &	
$14.8^{+60.1}_{-12.5}$ &	
$1.6^{+0.0}_{-0.0}$ &	
$-6.6^{+0.2}_{-0.2}$ &	
$27.8$ 	\\
L13 &
2016-01-21 &
5.30 (fixed) &	
3500 (fixed) &	
$27.4^{+0.8}_{-0.8}$ &	
1500 (fixed) &	
$86.5^{+12.0}_{-28.7}$ &	
$1.4^{+0.0}_{-0.0}$ &	
$-5.5^{+0.0}_{-0.0}$ &	
$2390$ 	\\
L14 &
2011-07-28 &
5.30 (fixed) &	
$26050^{+18800}_{-15790}$ &	
$0.6^{+2.2}_{-0.3}$ &	
1500 (fixed) &	
$16.7^{+64.4}_{-14.4}$ &	
$2.5^{+0.1}_{-0.2}$ &	
$-6.5^{+0.4}_{-0.2}$ &	
$1.4$ 	\\
L15 &
2012-03-16 &
5.30 (fixed) &	
$33500^{+12990}_{-14490}$ &	
$0.2^{+0.1}_{-0.1}$ &	
1500 (fixed) &	
$28.4^{+61.3}_{-25.6}$ &	
$2.4^{+0.0}_{-0.1}$ &	
$-7.0^{+0.2}_{-0.1}$ &	
$3.0$ 	\\
L16 &
2016-01-21 &
5.30 (fixed) &	
$48580^{+380}_{-920}$ &	
$0.0^{+0.0}_{-0.0}$ &	
1500 (fixed) &	
$37.2^{+30.0}_{-28.2}$ &	
$2.1^{+0.0}_{-0.0}$ &	
$-7.7^{+0.1}_{-0.1}$ &	
$168.6$ 	\\
\hline
\multicolumn{9}{c}{with dL/dt constraints} \\
L17 &
2011-07-28 &
$6.59^{+1.5}_{-0.91}$ &	
$6840^{+14190}_{-3210}$ &	
$31.8^{+6.7}_{-8.0}$ &	
$260^{+130}_{-1140}$ &	
2 (fixed) &	
$4.3^{+1.4}_{-1.1}$ &	
$\hphantom{-}1.6^{+2.9}_{-2.2}$ &	
$4.7$ 	\\
L18 &
2012-03-16 &
$4.83^{+0.22}_{-0.12}$ &	
$19120^{+25970}_{-15220}$ &	
$16.3^{+7.0}_{-6.1}$ &	
$1400^{+490}_{-580}$ &	
2 (fixed) &	
$2.2^{+0.6}_{-0.5}$ &	
$-2.8^{+1.1}_{-0.8}$ &	
$4.3$ 	\\
L19 &
2016-01-21 &
$4.50^{+0.11}_{-0.07}$ &	
$8860^{+1590}_{-1000}$ &	
$12.6^{+1.3}_{-1.3}$ &	
$1100^{+140}_{-220}$ &	
2 (fixed) &	
$1.7^{+0.2}_{-0.1}$ &	
$-3.1^{+0.5}_{-0.3}$ &	
$26.8$ 	\\
L20 &
2011-07-28 &
5.30 (fixed) &	
3500 (fixed) &	
$21.0^{+3.3}_{-2.2}$ &	
$580^{+40}_{-40}$ &	
2 (fixed) &	
$2.7^{+0.1}_{-0.1}$ &	
$-1.8^{+0.2}_{-0.1}$ &	
$11.7$ 	\\
L21 &
2012-03-16 &
5.30 (fixed) &	
3500 (fixed) &	
$19.4^{+3.8}_{-2.7}$ &	
$480^{+30}_{-30}$ &	
2 (fixed) &	
$2.4^{+0.1}_{-0.1}$ &	
$-1.5^{+0.2}_{-0.2}$ &	
$3.8$ 	\\
L22 &
2016-01-21 &
5.30 (fixed) &	
3500 (fixed) &	
$24.6^{+0.7}_{-0.7}$ &	
$340^{+30}_{-30}$ &	
2 (fixed) &	
$2.7^{+0.1}_{-0.1}$ &	
$-0.8^{+0.1}_{-0.1}$ &	
$1480$ 	\\
L23 &
2016-01-21 &
5.30 (fixed) &	
$14110^{+790}_{-740}$ &	
$15.1^{+0.8}_{-0.8}$ &	
$550^{+20}_{-20}$ &	
2 (fixed) &	
$2.8^{+0.0}_{-0.0}$ &	
$-0.8^{+0.1}_{-0.0}$ &	
$39.7$ 	\\
\hline
\multicolumn{9}{c}{near-IR only} \\
L24 &
2016-01-21 &
5.30 (fixed) &	
$15200^{+2890}_{-1810}$ &	
$36.7^{+1.8}_{-1.8}$ &	
1500 (fixed) &	
$50.2^{+19.7}_{-23.4}$ &	
$2.0^{+0.0}_{-0.0}$ &	
$-4.8^{+0.0}_{-0.0}$ &	
$264$ 	\\
\hline
\multicolumn{9}{c}{near-IR only with dL/dt} \\
L25 &
2016-01-21 &
5.30 (fixed) &	
$10880^{+1260}_{-1150}$ &	
$23.8^{+3.5}_{-3.3}$ &	
$170^{+300}_{-130}$ &	
2 (fixed) &	
$3.7^{+1.6}_{-0.9}$ &	
$1.3^{+3.2}_{-1.8}$ &	
$0.7$ 	\\
L26 &
2016-01-21 &
5.30 (fixed) &	
3500 (fixed) &	
$46.2^{+1.3}_{-1.4}$ &	
$140^{+190}_{-90}$ &	
2 (fixed) &	
$3.6^{+1.2}_{-0.8}$ &	
$1.2^{+2.4}_{-1.6}$ &	
$256$ 	\\
\hline
\multicolumn{9}{c}{near-IR only with dL/dt and $v_{\mathrm{ej}}=300^{+300}_{-150}$ prior} \\
L27 &
2016-01-21 &
5.30 (fixed) &	
$10860^{+1310}_{-1150}$ &	
$23.8^{+3.4}_{-3.4}$ &	
$430^{+130}_{-130}$ &	
2 (fixed) &	
$2.9^{+0.3}_{-0.2}$ &	
$-0.4^{+0.6}_{-0.4}$ &	
$2.6$ 	\\
\hline
\multicolumn{9}{c}{near-IR only with dL/dt and $v_{\mathrm{ej}}=300^{+30}_{-30}$ prior} \\
L28 &
2011-07-28 &
$5.13^{+0.14}_{-0.13}$ &	
3500 (fixed) &	
$18.3^{+11.2}_{-4.4}$ &	
$747^{+138}_{-119}$ &	
2 (fixed) &	
$2.4^{+0.2}_{-0.2}$ &	
$-2.4^{+0.4}_{-0.4}$ &	
$4.4$ 	\\
L29 &
2012-03-16 &
$5.22^{+0.13}_{-0.17}$ &	
3500 (fixed) &	
$11.1^{+9.9}_{-5.9}$ &	
$590^{+110}_{-90}$ &	
2 (fixed) &	
$2.5^{+0.2}_{-0.2}$ &	
$-2.3^{+0.5}_{-0.5}$ &	
$0.0$ 	\\
L30 &
2016-01-21 &
5.30 (fixed) &	
$10660^{+1270}_{-1120}$ &	
$24.4^{+3.4}_{-3.4}$ &	
$560^{+90}_{-80}$ &	
2 (fixed) &	
$2.7^{+0.1}_{-0.1}$ &	
$-0.8^{+0.2}_{-0.2}$ &	
$7.8$ 	\\
L31 & 
2016-01-21 &
$3.82^{+0.15}_{-0.13}$ &	
3500 (fixed) &	
$11.1^{+4.3}_{-4.2}$ &	
$170^{+40}_{-30}$ &	
2 (fixed) &	
$2.5^{+0.2}_{-0.2}$ &	
$-1.6^{+0.4}_{-0.4}$ &	
$0.2$ 	\\
\hline
\multicolumn{9}{c}{near-IR only with dL/dt and extrapolation to 2011-07-28} \\
L32 &
2016-01-21 &
$3.74^{+0.02}_{-0.02}$ &	
$4610^{+120}_{-120}$ &	
$0.0^{+0.1}_{-0.0}$ &	
$510^{+860}_{-450}$ &	
2 (fixed) &	
$2.0^{+1.9}_{-1.2}$ &	
$-5.1^{+3.9}_{-2.7}$ &	
$1600$ 	\\
L33 &
2016-01-21 &
$3.74^{+0.02}_{-0.01}$ &	
$4610^{+110}_{-110}$ &	
$0.0^{+0.1}_{-0.0}$ &	
$140^{+820}_{-90}$ &	
$4.9^{+79.1}_{-3.6}$ &	
$3.1^{+1.2}_{-1.9}$ &	
$-6.7^{+1.4}_{-2.2}$ &	
$1600$ 	\\
\hline
\multicolumn{9}{c}{near-IR only with dL/dt, $v_{\mathrm{ej}}=300^{+30}_{-30}$ prior, and 2011-07-28, 2012-03-16, 2016-01-21 extrapolation} \\
L34 &
2011-07-28 &
$5.54^{+0.08}_{-0.13}$ &
3500 (fixed) &
$54.0^{+10.8}_{-9.5}$ &
$1140^{+80}_{-80}$ &
2 (fixed) &
$2.5^{+0.2}_{-0.1}$ &
$-1.8^{+0.3}_{-0.3}$ &
$12.3$ \\
 &
2012-03-16 &
$4.54^{+0.04}_{-0.03}$ &
 &
$33.4^{+6.6}_{-5.9}$ &
$660^{+70}_{-50}$ &
 &
 &
 &
 \\
 &
2016-01-21 &
$4.10^{+0.11}_{-0.10}$ &
 &
$6.4^{+1.3}_{-1.1}$ &
$350^{+50}_{-40}$ &
 &
 &
 &
 \\
\hline
\hline
\end{tabular}
\end{minipage}
\end{center}
Constraints from MCMC models of the late-time SED.  The columns are the same as in Table \ref{tab:outburstmodels}, but with the addition of $v_{\mathrm{ej}}$, the velocity required for material ejected on 25 March 2009 to reach the inner edge of the dust shell.
\end{table*}

\begin{figure}
                \includegraphics[width=8.6cm, angle=0]{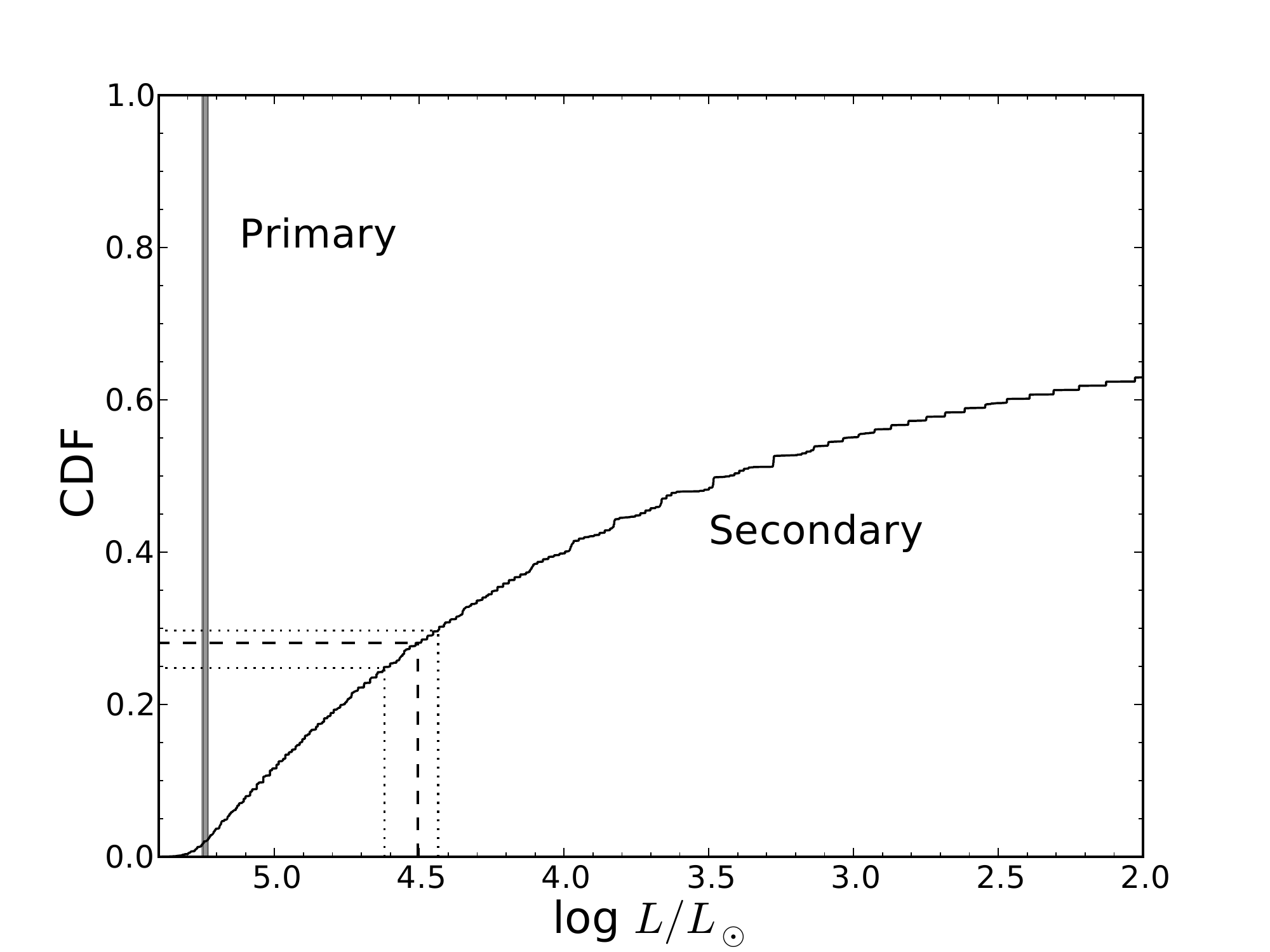}
		\caption{Cumulative distribution function for the luminosity of a surviving binary companion given our constraints on the progenitor mass (see Fig. \ref{fig:mass_pdf}).  
					The dashed line indicates the median luminosity from the MCMC SED modeling of the late-time source and the corresponding likelihood of a surviving secondary being this luminous.  Similarly, the dotted lines indicate the 10th and 90th percentile luminosities from the SED modeling.  The thin vertical band shows the constraint on the progenitor luminosity. \label{fig:secondarylum}}
					\end{figure}

		First, we consider whether the late-time emission could be due to a surviving binary companion obscured by dusty ejecta from the failed SN of the progenitor.  Massive stars have a large multiplicity fraction ($>82\%$; \citealt{Chini12}; \citealt{Sana12}).  We estimated the luminosity distribution of surviving companions of core-collapse SNe assuming passively evolving (i.e. no interactions) binaries with the progenitor masses drawn from the distribution in Fig. \ref{fig:progmass}.  We assumed a binary fraction of unity and a uniform secondary mass distribution following the simple models of \citet{Kochanek09}.  The secondary luminosity was the luminosity for a star with the mass of the secondary on the {\sc parsec} isochrone corresponding to the lifetime of the primary.
This yields the cumulative probability function of a surviving secondary brighter than a given luminosity shown in Fig. \ref{fig:secondarylum}.  If the binary fraction is unity, then there is a likelihood of roughly $30\%$ that the progenitor had a secondary at least as bright as the best estimate of log $L/L_{\odot}=4.50^{+0.10}_{-0.07}$ for the remaining luminosity in January 2016 (model L19; $90\%$ confidence intervals).  
		The likelihood drops in proportion to any reduction in the binary fraction.  However, a surviving secondary would not explain the continued late-time decay of the bolometric luminosity implied by the mid-IR devay.  The envelope of the secondary could have been shock heated by ejecta from the primary, making it overluminous, but the luminosity would only decay on a thermal time scale \citep[$>10^{3}$ yr;][]{Pan13,Shappee13}.

\begin{figure}
                                                             \includegraphics[width=8.6cm, angle=0]{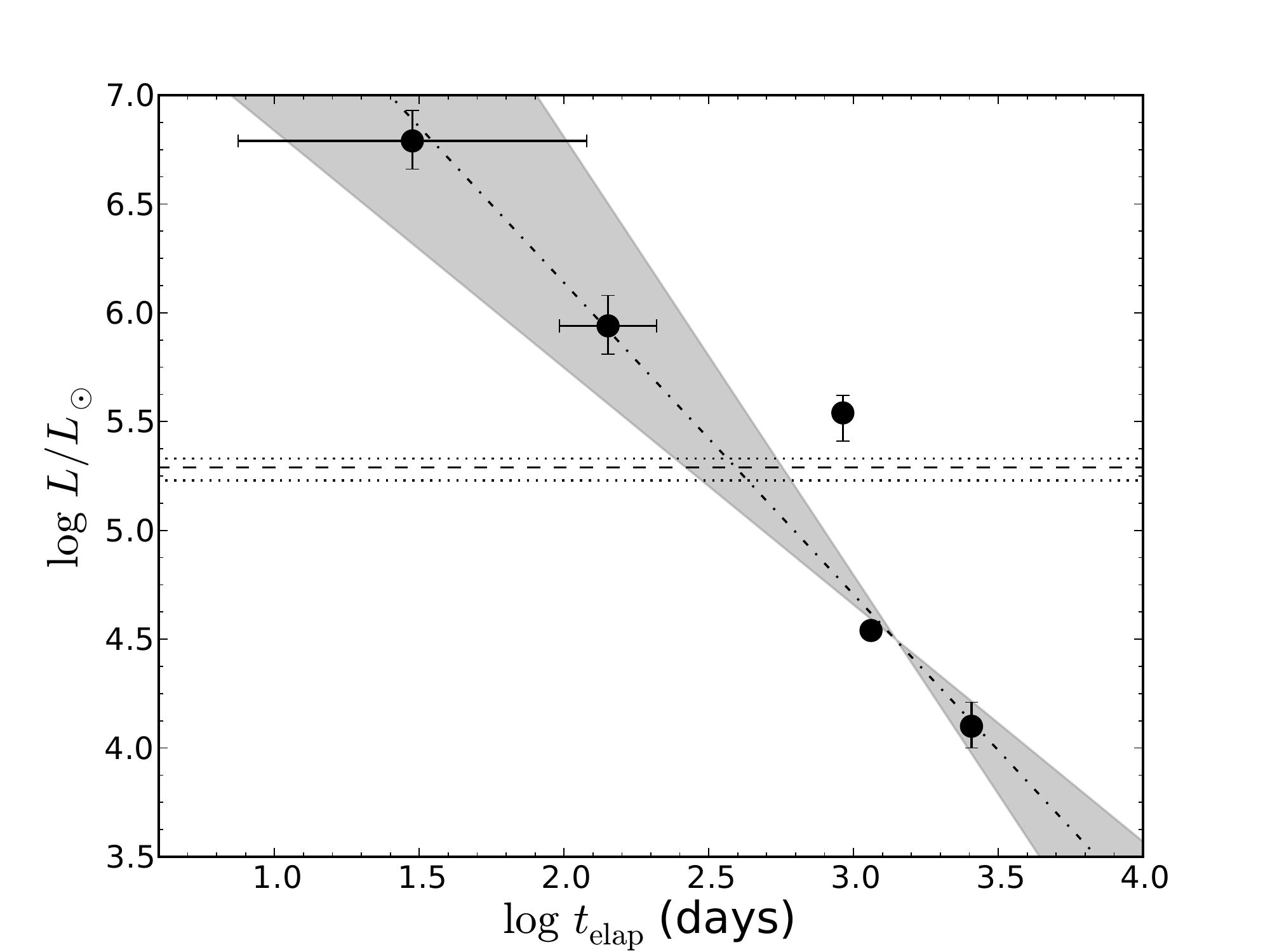}
							     \caption{Evolution of the bolometric luminosity as a function of elapsed time since the start of the optical outburst (between 25 November 2008 and 17 March 2009) as inferred from the SED model.  The vertical error bars on the points represent the $90\%$ confidence intervals from the MCMC modeling assuming $T_{*}=3500$ K (models O5 and O6) and that at late times a dusty shell is expanding self-consistently with the inferred velocity constraints (model L34).  The shaded gray region is defined by the bounds on the slope of the best-fit linear regressions for the luminosity given the uncertainty in the start of the transient.  The dot-dashed line has the slope of $-4/3$ expected for late-time fallback accretion disks \citep{Perna14}.  For reference, the horizontal dashed and dotted lines give the progenitor luminosity (on 2007-07-08 assuming $E(B$$-$$V)=0$; see P5 in Table \ref{tab:progmodels}) and its $90\%$ confidence interval. \label{fig:lum_evol}}
							     \end{figure}

					The late-time emission could be powered by fallback accretion onto a newly-formed black hole.  
					Although fallback is most commonly discussed as a possible power source for GRBs and other short-duration transients \citep{Woosley93,MacFadyen99,Kashiyama15}, fallback accretion may emit significant luminosity on much longer timescales ($>$yr) if the progenitor had sufficient angular momentum for material to circularize beyond the newly-formed black hole's innermost stable circular orbit.  Theoretical models have not been carried out for fallback accretion in the \citet{Lovegrove13} failed supernova scenario, but \citet{Perna14} find that for SN explosion energies leading to black hole remnants there is always a region of parameter space that leads to long-lived accretion disks.  Given sufficient angular momentum, super-Eddington accretion may be maintained for $\sim$10 yr with the accretion rate, $\dot{M}_{d}$, predicted to be roughly $\propto t^{-4/3}$ \citep{Perna14}.  After the accretion becomes sub-Eddington, the accretion rate declines more slowly with $\dot{M}_{d}\propto t^{-19/16}$ \citep{Cannizzo90}.
							     The evolution of the bolometric luminosity of N6946-BH1 may be well-fit by a power-law with a slope similar to these expectations for late-time fallback disks.  For example, the inferred luminosity evolution implied by models O5, O6, and L34 for a transient constrained to have begun between November 2008 and March 2009 is
							     $L\propto t^{-1.4\pm0.5}$
							     (see Fig. \ref{fig:lum_evol}).  
The ejecta from failed SNe, with their low velocities and relatively high densities, should be very efficient at forming dust \citep{Kochanek14c}.
							     Dust in marginally bound or slowly ejected material, could absorb the optical (and UV) flux emanating from a hot accretion disk and re-radiate it at the longer wavelengths shown to be slowly fading by the \emph{SST} data.  Although we are adopting a stellar spectrum as an input to our SED models, the details of the temperature and spectral characteristics of the intrinsic spectrum are unimportant when the optical depth is high.  We do, however, caution that the inferred slope for the bolometric evolution is model-dependent and less robust than the conclusion that by 2016 the bolometric luminosity had faded to significantly below that of the progenitor.

			\section{Alternatives}
			\label{sec:alternatives}

			We next consider alternative mechanisms for making a luminous star appear to vanish.  Could the failed SN candidate be an exotic type of variable star?
			The optical luminosities of Mira variables can decline by up to 8 mag over periods of hundreds of days \citep{Merrill40}.  As these pulsating stars cool, molecules such as titanium oxide can form high in the stellar atmosphere, reradiating the optical emission in the near-IR \citep{Reid02}.  The optical diminution of Miras is not preceded by an outburst, like that of N6946-BH1.  Moreover, N6946-BH1 has been optically faint for 7 years---much longer than the periods of Miras---and Miras do not experience changes in their bolometric luminosities as large as observed for N6946-BH1. 

			R Coronae Borealis (R Cor Bor) stars are a rare class of cool, carbon-rich, hydrogen-poor supergiants that repeatedly, but irregularly, decline for up to 8 magnitudes for hundreds to thousands of days, due to dust forming in the atmosphere \citep{Okeefe39,Clayton12}.  The absolute magnitudes of R Cor Bor stars range from $M_{V}\sim-5.2$ to $\sim-3.4$ at maximum \citep{Tisserand09}, significantly fainter than the progenitor of N6946-BH1 ($M_{V}\simeq-6.8$).  Since R Cor Bor stars are thought to arise from a very late final shell flash in AGB stars or the merger of two white dwarfs \citep{Iben96,Webbink84,Clayton12} there is no reason to suspect that a similar phenomenon occurs at the higher luminosity of N6946-BH1, and R Cor Bor stars do not experience optical outbursts like that seen for N6946-BH1 prior to its disappearance.

			An eclipsing binary with a long period, such as the recent one presented by \citet{Rodriguez16}, could possibly mimic the optical disappearance of N6946-BH1, but would not explain the optical outburst or the more gradual evolution of the IR luminosity.

			Could N6946-BH1 instead be a SN `impostor'?  At $M_{R}\lesssim-10.64$, the peak luminosity of the outburst likely lies within the faint end of events frequently considered to be non-terminal outbursts of luminous blue variables \citep{Smith11}.  The constraints on the ejecta velocity and transient duration are also consistent with the broad diversity of SN impostor properties.  The driving mechanism behind eruptive mass ejections in these massive stars is unclear.  There is significant evidence that some of these events may, in fact, be low-energy SNe \citep{Adams15,Adams16}.  The main evidence disfavoring N6946-BH1 being a SN `impostor' is that after eruptive mass loss a star is expected to be overluminous, not subluminous \citep{Kashi16}.  The conclusion from \S\ref{sec:latetime} is that N6946-BH1 has faded below the luminosity of its progenitor, which makes it unlikely that the star survived.

Although we only consider the simple case of obscuration of spherically symmetric dust, we expect this basic conclusion to largely be valid for shells with inhomegenities and some asymetries \citep[see discussion in][]{Kochanek12}. 
For example, \citet{Kashi17} propose a complex dust geometry to evade the conclusions
of our models with a spherical dust distribution.  In essence, they propose a
toroidal dust distribution viewed edge with the energy escaping along the
axis of the torus without significant heating the torus.  We used {\sc dusty} to explore the
limiting case of this geometry, a slab of dust illuminated from one side and
observed from the other, and found that significant fractions of
the energy were still transmitted through the slab and visible as hot dust emission
even when the visual opacity of the slab was very large (25\% of the energy for $\tau_V=100$).
In an actual torus, this fraction will increase because energy from heating
the inner edge of the torus which is not transmitted will also contribute to
the heating of other areas of the torus instead of simply escaping. In the
vertical direction, the torus also has to have negligible optical depth because
for a low optical depth the star will be visible in scattered light and for higher
optical depth the geometry returns to being effectively spherical.

			Our primary concern with the failed SN interpretation is that it does not provide a natural explanation for the rising IR flux in the final years prior to the optical outburst.  
        Various mechanisms have been proposed to enhance mass loss in the final years before a SN (e.g., \citealt{Shiode14}; \citealt{Mcley14}), though mainly in the context of explaining precursor eruptions.  Dramatic eruptions must be relatively rare ($\sim10\%$) given the relatively low frequency of Type IIn SNe.  Even if precursor variability is a continuum, significantly enhanced mass loss is likely not a generic outcome given that in the few cases where there are multi-epoch observations of the SN progenitor little, if any, variability is observed \citep[see][]{Kochanek16}.

A stellar merger could explain many of the properties of N6946-BH1, as Roche lobe overflow could give rise to increasing IR and decreasing optical flux prior to an optical outburst triggered by contact \citep{Pejcha14,Pejcha16,MacLeod17}.  The subsequent inspiral of the secondary through the envelope could lead to significant mass loss and dust formation \citep{MacLeod17}.  However, in such a scenario the mass outflow would likely be on order the escape velocity of the primary \citep{Ivanova13,Pejcha16b}, which, for the RSG progenitor of N6946-BH1 would be somewhat lower than the velocity constraint we found in \S\ref{sec:outburst} (85-115 $\mathrm{km}$ $\mathrm{s}^{-1}$ vs 170-560 $\mathrm{km}$ $\mathrm{s}^{-1}$).
  Also, the merger remnant would be expected to be significantly more luminous than the progenitor for $\sim$a thermal time \citep{Ivanova13}, as appears to be the case for the well-studied Galactic stellar mergers V838 Mon \citep{Tylenda05a}, V4332 Sgr \citep{Tylenda05b}, and V1309 Scorpii \citep{Tylenda16}, as well as for the recent proposed massive star mergers in NGC 4490 \citep{Smith16} and M101 \citep{Blagorodnova17}.  

	\section{Summary and Conclusions}
	\label{sec:summary}

			We present new late-time \emph{HST}, \emph{SST}, and LBT observations of the failed supernova candidate found by \citet{Gerke15}.  We also analyze archival imaging of the progenitor and find that between 1999 and 2009 the progenitor maintained a roughly constant bolometric luminosity while fading in the optical and brightening in the mid-IR.  The progenitor SED can be well-fit by a $\sim$$25$ $M_{\odot}$ RSG.  This lies directly within the mass range of the missing RSG supernova progenitors and high core ``compactness" identified in theoretical models as most likely to give rise to failed supernovae and direct black hole formation.  
			In 2009 the progenitor underwent a weak ($\sim$ few $10^{6}~L_{\odot}$), but long (3--11 month), optical outburst.  The transient is broadly consistent with the failed SN models of \citet{Lovegrove13}, although the data on the optical transient are limited.  Between 3 and 11 months from the start of the optical outburst the optical flux collapsed to be far below that of the progenitor, though the source faded more slowly in the mid-IR.  This suggests that dust formed in material ejected during the outburst with $170\gtrsim v_{\mathrm{ej}}< 560~\mathrm{km}\>\mathrm{s}^{-1}$.  These constraints on the peak luminosity, transient duration, and ejecta velocity are consistent with numerical simulations of failed SNe \citep{Lovegrove13}, but they also may be consistent with other phenomenon such a SN impostor or stellar merger.

			Our late-time data shows that the source is now $>5$ mag fainter than the progenitor in the optical and, though still fading, has returned to the pre-transient flux in the mid-IR.  The bolometric luminosity of the source is fading with 
			$L\propto t^{-1.4\pm0.5}$
			, which is consistent with the models of late-time fallback accretion onto a black hole \citep{Perna14}.  The bolometric luminosity of N6946-BH1 is now significantly fainter than the progenitor, suggesting that the progenitor did not survive.  We propose that the late-time detection of near-IR emission may be due to fallback accretion onto a newly formed black hole obscured by dust that formed in a weakly-ejected envelope.

			N6946-BH1 merits further study.  It would be valuable to constrain the luminosity variability of the progenitor at even earlier times with other archival data.  New observations are also needed to confirm that N6946-BH1 is a failed SN.  If the late-time luminosity is powered by fallback accretion, it may be possible to detect X-rays with the Chandra X-ray Observatory as long as the neutral hydrogen column depth is not too large.  
			If X-rays are not detected it will be especially important to continue to monitor N6946-BH1 in the optical to make sure that the source does not start to rebrighten due to a decreasing optical depth and in the mid-IR to see if the $3.6~\mu\mathrm{m}$ and $4.5~\mathrm{m}$ fluxes continue to decrease.
Ultimately, observations at $10-20~\mu\mathrm{m}$ with the \emph{James Webb Space Telescope} may be needed to verify that a surviving star is not hidden by cooler dust than can be probed with \emph{SST}.  If confirmed, N6946-BH1 would be the first failed SN and first black hole birth ever discovered and would resolve the problem of the missing high-mass SN progenitors.

\section*{Acknowledgements}
We thank Ron Arbour for sharing his images of NGC 6946.  We thank Klaas Wiersema and Andrew Levan for identifying observations covering N6946-BH1 in the INT archive.  We thank John Beacom and the anonymous referee for helpful comments on the manuscript.
Financial support for this work was provided by the NSF through grant AST-1515876.
 This work is based in part on observations made with the Spitzer Space Telescope, which is operated by the Jet Propulsion Laboratory, California Institute of Technology under a contract with NASA, and in part on observations made with the NASA/ESA \emph{Hubble Space Telescope} obtained at the Space Telescope Institute, which is operated by the Association of Universities for Research in Astronomy, Inc., under NASA contract NAS 5-26555. These observations are associated with program GO-14266.
This work is based in part on observations made with the Large
Binocular Telescope. 
The LBT is an international collaboration among institutions in the United States, Italy and Germany. The LBT Corporation partners are: The University of Arizona on behalf of the Arizona university system; Istituto Nazionale di Astrofisica, Italy;  LBT Beteiligungsgesellschaft, Germany, representing the Max Planck Society, the Astrophysical Institute Potsdam, and Heidelberg University; The Ohio State University; The Research Corporation, on behalf of The University of Notre Dame, University of Minnesota and University of Virginia.
This research used the facilities of the Canadian Astronomy Data Centre operated by the National Research Council of Canada with the support of the Canadian Space Agency. 

\bibliography{references}
\bibliographystyle{mn2e}
\clearpage

\begin{table}
\caption{Photometry}
\begin{tabular}{cccc}
\hline
\hline
{MJD} & {Filter} & {Magnitude} & {Telescope} \\
\hline
51341.2 &	$B$	&	$23.3\pm0.1$	&	INT	\\
51341.2	&	$r$	&	$22.0\pm0.1$	&	INT	\\
51341.2 &	$i$	&	$21.4\pm0.1$	&	INT	\\
52498.9 &	$V$	&	$22.3\pm0.1$	&	INT	\\ 
52498.9	&	$i$	&	$21.2\pm0.1$	&	INT	\\
52935.0	&	$R$	&	$21.5\pm0.3$	&	CFHT	\\
52935.0	&	$V$	&	$21.7\pm0.2$	&	CFHT	\\
53166.8	 & 	$3.6\>\mu\mathrm{m}$	 & 	$17.84\pm0.08$	 & 	\emph{SST}	\\
53166.8	 & 	$4.5\>\mu\mathrm{m}$	 & 	$17.69\pm0.09$	 & 	\emph{SST}	\\
53260.3	 & 	$4.5\>\mu\mathrm{m}$	 & 	$17.24\pm0.08$	 & 	\emph{SST}	\\
53334.7	 & 	$3.6\>\mu\mathrm{m}$	 & 	$17.42\pm0.05$	 & 	\emph{SST}	\\
53334.7	 & 	$4.5\>\mu\mathrm{m}$	 & 	$17.24\pm0.07$	 & 	\emph{SST}	\\
53553.0 &	$B$	&	$22.8\pm0.1$	&	CFHT	\\
53553.0	&	$R$	&	$21.3\pm0.3$	&	CFHT	\\
53553.0	&	$U$	&	$23.5\pm0.2$	&	CFHT	\\
53553.0	&	$V$	&	$21.7\pm0.1$	&	CFHT	\\
53571.2	 & 	$4.5\>\mu\mathrm{m}$	 & 	$17.32\pm0.06$	 & 	\emph{SST}	\\
53571.2	 & 	$3.6\>\mu\mathrm{m}$	 & 	$17.49\pm0.06$	 & 	\emph{SST}	\\
53571.2	 & 	$4.5\>\mu\mathrm{m}$	 & 	$17.35\pm0.07$	 & 	\emph{SST}	\\
53630.8	 & 	$4.5\>\mu\mathrm{m}$	 & 	$16.46\pm0.11$	 & 	\emph{SST}	\\
53676.0	 & 	$3.6\>\mu\mathrm{m}$	 & 	$17.39\pm0.05$	 & 	\emph{SST}	\\
53676.0	 & 	$4.5\>\mu\mathrm{m}$	 & 	$17.12\pm0.05$	 & 	\emph{SST}	\\
53959.5	 & 	$4.5\>\mu\mathrm{m}$	 & 	$16.98\pm0.07$	 & 	\emph{SST}	\\
54006.2	 & 	$3.6\>\mu\mathrm{m}$	 & 	$17.04\pm0.04$	 & 	\emph{SST}	\\
54006.2	 & 	$4.5\>\mu\mathrm{m}$	 & 	$16.92\pm0.05$	 & 	\emph{SST}	\\
54065.9	 & 	$3.6\>\mu\mathrm{m}$	 & 	$17.17\pm0.05$	 & 	\emph{SST}	\\
54065.9	 & 	$4.5\>\mu\mathrm{m}$	 & 	$16.90\pm0.05$	 & 	\emph{SST}	\\
54098.0	 & 	$3.6\>\mu\mathrm{m}$	 & 	$17.02\pm0.04$	 & 	\emph{SST}	\\
54098.0	 & 	$4.5\>\mu\mathrm{m}$	 & 	$16.85\pm0.052$	 & 	\emph{SST}	\\
54285.0	 & 	$3.6\>\mu\mathrm{m}$	 & 	$17.14\pm0.05$	 & 	\emph{SST}	\\
54285.0	 & 	$4.5\>\mu\mathrm{m}$	 & 	$16.94\pm0.05$	 & 	\emph{SST}	\\
54289.2	 & 	$F606W$	 & 	$23.09\pm0.01$	 & 	\emph{HST}	\\
54289.3	 & 	$F814W$	 & 	$20.77\pm0.01$	 & 	\emph{HST}	\\
54324.4	 & 	$3.6\>\mu\mathrm{m}$	 & 	$17.19\pm0.05$	 & 	\emph{SST}	\\
54324.4	 & 	$4.5\>\mu\mathrm{m}$	 & 	$16.97\pm0.05$	 & 	\emph{SST}	\\
54395.9	 & 	$3.6\>\mu\mathrm{m}$	 & 	$16.96\pm0.04$	 & 	\emph{SST}	\\
54461.1	 & 	$3.6\>\mu\mathrm{m}$	 & 	$16.86\pm0.03$	 & 	\emph{SST}	\\
54461.1	 & 	$4.5\>\mu\mathrm{m}$	 & 	$16.66\pm0.04$	 & 	\emph{SST}	\\
54492.9	 & 	$3.6\>\mu\mathrm{m}$	 & 	$16.93\pm0.04$	 & 	\emph{SST}	\\
54589.4      &       $B$     &       $24.43\pm0.04$  &       LBT     \\
54589.4      &       $R$     &       $21.70\pm0.01$  &       LBT     \\
54589.5      &       $U$     &       $25.18\pm0.26$  &       LBT     \\
54590.4      &       $R$     &       $21.73\pm0.01$  &       LBT     \\
54652.4      &       $B$     &       $24.41\pm0.03$  &       LBT     \\
54652.4      &       $U$     &       $27.39\pm1.90$  &       LBT     \\
54652.4      &       $V$     &       $23.01\pm0.01$  &       LBT     \\
54665.8	 & 	$3.6\>\mu\mathrm{m}$	 & 	$16.700\pm0.037$	 & 	\emph{SST}	\\
54665.8	 & 	$4.5\>\mu\mathrm{m}$	 & 	$16.541\pm0.041$	 & 	\emph{SST}	\\
54761.8     &      $V$      &      $>18.8$  &      Arbour  \\
54763.8     &      $V$      &      $>18.6$  &      Arbour  \\
54795.1      &       $B$     &       $23.32\pm0.01$  &       LBT     \\
54795.1      &       $V$     &       $21.27\pm0.01$  &       LBT     \\
54806.9     &      $V$      &      $>18.7$  &      Arbour  \\
54811.8     &      $V$      &      $>18.3$  &      Arbour  \\
54829.8     &      $V$      &      $>18.7$  &      Arbour  \\
54836.8     &      $V$      &      $>17.5$  &      Arbour  \\
54837.8     &      $V$      &      $>18.7$  &      Arbour  \\
54844.8     &      $V$      &      $>17.5$  &      Arbour  \\
54854.8     &      $V$      &      $>17.5$  &      Arbour  \\
54907.5	 & 	$R$	 & 	$19.02\pm0.07$	 & 	PTF	\\
54907.5	 & 	$R$	 & 	$19.00\pm0.07$	 & 	PTF	\\
54915.5      &       $B$     &       $22.14\pm0.01$  &       LBT     \\
54915.5      &       $R$     &       $18.98\pm0.01$  &       LBT     \\
54916.5	 & 	$g$	 & 	$>19.6$	 & 	PTF	\\
\end{tabular}
\label{tab:photometry}
\end{table}

\begin{table}
\contcaption{}
\begin{tabular}{cccc}
\hline
\hline
{MJD} & {Filter} & {Magnitude} & {Telescope} \\
\hline
54916.5  &      $g$      &      $19.74\pm0.20$   &      PTF     \\
54918.5  &      $g$      &      $19.82\pm0.13$   &      PTF     \\
54922.5  &      $g$      &      $19.63\pm0.14$   &      PTF     \\
54922.5  &      $g$      &      $>20.7$  &      PTF     \\
54977.5  &      $R$      &      $19.57\pm0.11$   &      PTF     \\
54984.5	 & 	$R$	 & 	$>19.7$	 & 	PTF	\\
54984.5	 & 	$R$	 & 	$>18.9$	 & 	PTF	\\
55004.3      &       $R$     &       $21.03\pm0.01$  &       LBT     \\
55004.3      &       $U$     &       $24.66\pm0.15$  &       LBT     \\
55004.3      &       $V$     &       $22.59\pm0.01$  &       LBT     \\
55018.4	 & 	$R$	 & 	$>20.5$	 & 	PTF	\\
55020.4	 & 	$R$	 & 	$>20.2$	 & 	PTF	\\
55020.4	 & 	$R$	 & 	$>20.2$	 & 	PTF	\\
55022.4	 & 	$R$	 & 	$>20.8$	 & 	PTF	\\
55022.4	 & 	$R$	 & 	$>20.7$	 & 	PTF	\\
55025.4	 & 	$R$	 & 	$>21.1$	 & 	PTF	\\
55028.4	 & 	$R$	 & 	$>21.4$	 & 	PTF	\\
55034.4	 & 	$R$	 & 	$>21.3$	 & 	PTF	\\
55034.4	 & 	$R$	 & 	$>21.3$	 & 	PTF	\\
55042.3	 & 	$R$	 & 	$>21.7$	 & 	PTF	\\
55048.4	 & 	$R$	 & 	$>20.6$	 & 	PTF	\\
55094.3	 & 	$R$	 & 	$>21.2$	 & 	PTF	\\
55107.2	 & 	$R$	 & 	$>20.4$	 & 	PTF	\\
55107.3	 & 	$R$	 & 	$>20.3$	 & 	PTF	\\
55123.1	 & 	$R$	 & 	$>21.4$	 & 	PTF	\\
55123.2	 & 	$R$	 & 	$>21.4$	 & 	PTF	\\
55124.1      &       $B$     &       $25.70\pm0.01$  &       LBT     \\
55124.1      &       $U$     &       $24.83\pm0.19$  &       LBT     \\
55124.1      &       $V$     &       $25.61\pm0.13$  &       LBT     \\
55126.1      &       $B$     &       $25.68\pm0.12$  &       LBT     \\
55126.1      &       $U$     &       $24.33\pm0.13$  &       LBT     \\
55126.1      &       $V$     &       $26.12\pm0.21$  &       LBT     \\
55273.5      &       $B$     &       $25.51\pm0.15$  &       LBT     \\
55273.5      &       $R$     &       $24.45\pm0.12$  &       LBT     \\
55273.5      &       $U$     &       $24.78\pm0.30$  &       LBT     \\
55273.5      &       $V$     &       $>25.97$        &       LBT     \\
55536.1      &       $B$     &       $26.10\pm0.14$  &       LBT     \\
55536.1      &       $R$     &       $24.93\pm0.10$  &       LBT     \\
55543.2      &       $R$     &       $24.98\pm0.21$  &       LBT     \\
55543.2      &       $V$     &       $>26.44$        &       LBT     \\
55717.4      &       $B$     &       $25.84\pm0.12$  &       LBT     \\
55717.4      &       $U$     &       $26.16\pm0.58$  &       LBT     \\
55717.4      &       $V$     &       $26.47\pm0.27$  &       LBT     \\
55769.9	 & 	$3.6\>\mu\mathrm{m}$	 & 	$16.96\pm0.04$	 & 	\emph{SST}	\\
55769.9	 & 	$4.5\>\mu\mathrm{m}$	 & 	$16.34\pm0.03$	 & 	\emph{SST}	\\
55824.3      &       $B$     &       $25.58\pm0.13$  &       LBT     \\
55824.3      &       $R$     &       $24.74\pm0.09$  &       LBT     \\
55824.3      &       $V$     &       $25.45\pm0.12$  &       LBT     \\
55827.2      &       $B$     &       $25.72\pm0.10$  &       LBT     \\
55827.2      &       $U$     &       $24.34\pm0.15$  &       LBT     \\
55827.2      &       $R$     &       $24.63\pm0.07$  &       LBT     \\
55827.3      &       $V$     &       $26.03\pm0.21$  &       LBT     \\
55828.3      &       $B$     &       $25.62\pm0.09$  &       LBT     \\
55828.3      &       $R$     &       $24.81\pm0.09$  &       LBT     \\
55828.3      &       $V$     &       $25.88\pm0.14$  &       LBT     \\
55884.1      &       $R$     &       $24.88\pm0.15$  &       LBT     \\
55884.1      &       $B$     &       $25.89\pm0.18$  &       LBT     \\
55884.1      &       $V$     &       $27.10\pm0.66$  &       LBT     \\
56002.0	 & 	$3.6\>\mu\mathrm{m}$	 & 	$17.49\pm0.06$	 & 	\emph{SST}	\\
56002.0	 & 	$4.5\>\mu\mathrm{m}$	 & 	$16.68\pm0.04$	 & 	\emph{SST}	\\
56045.4      &       $B$     &       $25.69\pm0.13$  &       LBT     \\
56045.5      &       $R$     &       $24.74\pm0.13$  &       LBT     \\
56045.5      &       $U$     &       $25.28\pm0.33$  &       LBT     \\
56045.5      &       $V$     &       $26.63\pm0.35$  &       LBT     \\
\end{tabular}
\end{table}

\begin{table}
\contcaption{}
\begin{tabular}{cccc}
\hline
\hline
{MJD} & {Filter} & {Magnitude} & {Telescope} \\
\hline
56090.4      &       $B$     &       $25.48\pm0.11$  &       LBT     \\
56090.4      &       $R$     &       $24.78\pm0.09$  &       LBT     \\
56090.4      &       $U$     &       $26.29\pm0.72$  &       LBT     \\
56090.4      &       $V$     &       $26.04\pm0.20$  &       LBT     \\
56093.4      &       $B$     &       $25.53\pm0.11$  &       LBT     \\
56093.4      &       $R$     &       $24.77\pm0.09$  &       LBT     \\
56093.4      &       $U$     &       $25.64\pm0.43$  &       LBT     \\
56093.4      &       $V$     &       $25.98\pm0.16$  &       LBT     \\
56215.1      &       $B$     &       $25.47\pm0.12$  &       LBT     \\
56215.1      &       $U$     &       $24.92\pm0.27$  &       LBT     \\
56215.1      &       $V$     &       $27.16\pm0.62$  &       LBT     \\
56217.1      &       $B$     &       $26.08\pm0.21$  &       LBT     \\
56217.1      &       $R$     &       $24.77\pm0.15$  &       LBT     \\
56217.1      &       $U$     &       $>25.13$        &       LBT     \\
56217.2      &       $V$     &       $>26.55$        &       LBT     \\
56449.4      &       $B$     &       $25.61\pm0.12$  &       LBT     \\
56449.4      &       $R$     &       $24.82\pm0.10$  &       LBT     \\
56449.4      &       $U$     &       $25.41\pm0.43$  &       LBT     \\
56449.4      &       $V$     &       $26.58\pm0.47$  &       LBT     \\
56453.4      &       $B$     &       $25.73\pm0.11$  &       LBT     \\
56453.4      &       $R$     &       $24.79\pm0.08$  &       LBT     \\
56453.4      &       $U$     &       $26.63\pm1.07$  &       LBT     \\
56453.4      &       $V$     &       $25.93\pm0.20$  &       LBT     \\
56629.1      &       $B$     &       $25.67\pm0.14$  &       LBT     \\
56629.1      &       $R$     &       $25.00\pm0.16$  &       LBT     \\
56629.1      &       $U$     &       $25.46\pm0.43$  &       LBT     \\
56742.6	 & 	$3.6\>\mu\mathrm{m}$	 & 	$17.95\pm0.08$	 & 	\emph{SST}	\\
56812.4      &       $B$     &       $25.57\pm0.11$  &       LBT     \\
56812.4      &       $R$     &       $25.05\pm0.15$  &       LBT     \\
56812.4      &       $U$     &       $25.29\pm0.35$  &       LBT     \\
56812.4      &       $V$     &       $26.40\pm0.32$  &       LBT     \\
56815.4      &       $B$     &       $25.93\pm0.15$  &       LBT     \\
56815.4      &       $R$     &       $25.00\pm0.10$  &       LBT     \\
56815.4      &       $U$     &       $26.20\pm0.76$  &       LBT     \\
56815.4      &       $V$     &       $26.00\pm0.16$  &       LBT     \\
56833.4      &       $B$     &       $25.73\pm0.14$  &       LBT     \\
56833.4      &       $R$     &       $24.82\pm0.10$  &       LBT     \\
56833.4      &       $U$     &       $>25.24$        &       LBT     \\
56833.4      &       $V$     &       $25.73\pm0.13$  &       LBT     \\
56836.4      &       $B$     &       $25.63\pm0.11$  &       LBT     \\
56836.4      &       $R$     &       $24.92\pm0.09$  &       LBT     \\
56836.4      &       $U$     &       $26.35\pm0.75$  &       LBT     \\
56836.4      &       $V$     &       $26.30\pm0.23$  &       LBT     \\
56839.3      &       $R$     &       $24.67\pm0.09$  &       LBT     \\
56839.3      &       $U$     &       $24.83\pm0.23$  &       LBT     \\
56839.3      &       $V$     &       $27.29\pm0.64$  &       LBT     \\
56900.2      &       $B$     &       $25.70\pm0.01$  &       LBT     \\
56900.2      &       $V$     &       $26.44\pm0.27$  &       LBT     \\
56916.2	 & 	$4.5\>\mu\mathrm{m}$	 & 	$17.07\pm0.05$	 & 	\emph{SST}	\\
56925.2      &       $B$     &       $26.04\pm0.14$  &       LBT     \\
56925.2      &       $R$     &       $24.73\pm0.08$  &       LBT     \\
56925.2      &       $U$     &       $25.59\pm0.30$  &       LBT     \\
56925.2      &       $V$     &       $26.01\pm0.14$  &       LBT     \\
56945.6	 & 	$4.5\>\mu\mathrm{m}$	 & 	$17.26\pm0.06$	 & 	\emph{SST}	\\
56981.1      &       $B$     &       $25.89\pm0.12$  &       LBT     \\
56981.1      &       $R$     &       $24.67\pm0.09$  &       LBT     \\
56981.1      &       $U$     &       $27.07\pm1.17$  &       LBT     \\
56981.1      &       $V$     &       $26.12\pm0.20$  &       LBT     \\
57132.4      &       $B$     &       $25.71\pm0.11$  &       LBT     \\
57132.4      &       $U$     &       $25.56\pm0.34$  &       LBT     \\
57132.4      &       $R$     &       $24.58\pm0.08$  &       LBT     \\
57132.5      &       $V$     &       $26.08\pm0.21$  &       LBT     \\
57163.4      &       $B$     &       $26.66\pm0.44$  &       LBT     \\
57163.4      &       $R$     &       $24.92\pm0.19$  &       LBT     \\
\end{tabular}
\end{table}

\begin{table}
\contcaption{}
\begin{tabular}{cccc}
\hline
\hline
{MJD} & {Filter} & {Magnitude} & {Telescope} \\
\hline
57163.4      &       $V$     &       $>26.34$        &       LBT     \\
57281.9  &      $4.5\>\mu\mathrm{m}$     &      $17.50\pm0.08$   &      \emph{SST}      \\
57294.0  &      $4.5\>\mu\mathrm{m}$     &      $17.35\pm0.07$   &      \emph{SST}      \\
57303.2	 & 	UVIS $F606W$	 & 	$28.44\pm0.46$	 & 	\emph{HST}	\\
57303.3	 & 	UVIS $F814W$	 & 	$26.02\pm0.16$	 & 	\emph{HST}	\\
57303.3	 & 	IR $F110W$	 & 	$23.75\pm0.02$	 & 	\emph{HST}	\\
57303.3	 & 	IR $F160W$	 & 	$22.38\pm0.02$	 & 	\emph{HST}	\\
57309.1  &       $B$     &       $25.73\pm0.10$  &       LBT     \\
57309.1      &       $U$     &       $25.28\pm0.26$  &       LBT     \\
57309.1      &       $R$     &       $24.81\pm0.08$  &       LBT     \\
57309.1      &       $V$     &       $25.63\pm0.10$  &       LBT     \\
57322.1	 & 	$4.5\>\mu\mathrm{m}$	 & 	$17.39\pm0.05$	 & 	\emph{SST}	\\
57364.1      &       $B$     &       $25.56\pm0.08$  &       LBT     \\
57364.1      &       $U$     &       $25.33\pm0.25$  &       LBT     \\
57364.1      &       $R$     &       $24.84\pm0.11$  &       LBT     \\
57408.2	 & 	$3.6\>\mu\mathrm{m}$	 & 	$18.67\pm0.12$	 & 	\emph{SST}	\\
57408.2	 & 	$4.5\>\mu\mathrm{m}$	 & 	$17.42\pm0.06$	 & 	\emph{SST}	\\
\hline
\hline
\end{tabular}
\end{table}

\clearpage

\end{document}